\documentclass{UnderReview}

\usepackage[colorlinks,urlcolor=blue,linkcolor=blue,citecolor=blue]{hyperref}

\usepackage{color,array,amsthm}

\usepackage{graphicx}

\usepackage[utf8]{inputenc}
\usepackage{graphicx}
\graphicspath{{./fig/}}
\usepackage{caption}

\usepackage[ruled,vlined]{algorithm2e}
\LinesNumbered
\SetKw{Or}{or}

\usepackage{xcolor}
\usepackage{tcolorbox}
\usepackage{gnc}
\usepackage{amsmath}
\usepackage{amssymb}
\usepackage{longtable,tabularx}

\DeclareMathOperator*{\E}{\mathbb{E}}

\newcommand{\kyle}[1]{{\footnotesize \sf {\color{blue}{\bf Kyle:} #1}}}
\newcommand{\ani}[1]{{\footnotesize \sf {\color{red}{\bf Ani:} #1}}}
\newcommand{\dan}[1]{{\footnotesize \sf {\color{green!70!black}{\bf Dan:} #1}}}

\renewcommand{\kyle}[1]{}
\renewcommand{\ani}[1]{}
\renewcommand{\dan}[1]{}

\begin{document}
\newcommand{\Tr}{^\mathsf{T}}
\renewcommand{\(}{\left(}
\renewcommand{\)}{\right)}
\newcommand{\q}{\mathbf{q}}
\newcommand{\x}{\mathbf{x}}
\newcommand{\f}{\mathbf{f}}
\renewcommand{\u}{\mathbf{u}}
\newcommand{\w}{\mathbf{w}}
\newcommand{\nchoosek}[2]{\left(\begin{matrix}
		#1\\
		#2
	\end{matrix}\right)}

\newcommand{\updates}[1]{\textcolor{black}{#1}}

\title{Trajectory Planning with Deep Reinforcement Learning in High-Level Action Spaces}

\author{Kyle R. Williams, Rachel Schlossman, Daniel Whitten}
\author{Joe Ingram, Srideep Musuvathy, James Pagan, Kyle A. Williams}
\affil{Sandia National Laboratories, Albuquerque, NM} 

\author{Sam Green, Anirudh Patel}
\affil{Semiotic Labs, Los Altos, CA} 

\author{Anirban Mazumdar}
\affil{Georgia Institute of Technology, Atlanta, GA}

\author{Julie Parish}
\affil{Sandia National Laboratories, Albuquerque, NM} 



\authoraddress{Authors' addresses: Kyle R. Williams, Rachel Schlossman, Daniel Whitten, Joe Ingram, Srideep Musuvathy, James Pagan, Kyle A. Williams and Julie Parish are with Sandia National Laboratories, Albuquerque, NM, emails: \{kwilli2, rschlos, wdwhitt, jbingra, smusuva, jepagan, kwilli3, jparish\}@sandia.gov.  Sam Green and Anirudh Patel are with Semiotic Labs, Los Altos, CA (work was performed while at Sandia National Labs), email: \{sam, anirudh\}@semiotic.ai.  Anirban Mazumdar is with the Georgia Institute of Technology, Atlanta, GA, email: anirban.mazumdar@me.gatech.edu. {\itshape (Corresponding author: K.R.Williams)}.}


\markboth{AUTHOR ET AL.}{SHORT ARTICLE TITLE}
\maketitle

\begin{abstract} \updates{This paper presents a technique for trajectory planning based on parameterized high-level actions.  These high-level actions are sub trajectories that have variable shape and duration.  The use of high-level actions can improve the performance of guidance algorithms.  Specifically, we show how the use of high-level actions improves the performance of guidance policies that are generated via reinforcement learning (RL).  Reinforcement learning has shown great promise for solving complex control, guidance, and coordination problems but can still suffer from long training times and poor performance.  This work shows how the use of high-level actions reduces the required number of training steps and increases the path performance of an RL-trained guidance policy.  We demonstrate the method on a space-shuttle guidance example.  We show the proposed method increases the path performance (latitude range) by 18\% percent compared to a baseline RL implementation.  Similarly, we show the proposed method achieves steady state during training with approximately 75\% fewer training steps. We also show how the guidance policy enables effective performance in an obstacle field.  Finally, this paper develops a loss function term for policy-gradient-based Deep RL, which is analogous to an anti-windup mechanism in feedback control.  We demonstrate that the inclusion of this term in the underlying optimization increases the average policy return in our numerical example.}
\end{abstract}


\section{Introduction}
Autonomous vehicles have become an active area of research over the past several decades.  One of the biggest challenges for these vehicles is trajectory planning (also called guidance).  Trajectory planning or guidance involves solving a complex optimization problem in real-time, reliably, and with limited computational resources.  Many advances have been made in trajectory planning and optimization.  Examples include nonlinear programming methods \cite{betts2010practical}, sampling based methods \cite{lavalle2001randomized}, and discrete motion planning such as A* \cite{hart1968formal}, Hybrid A* \cite{dolgov2008practical}, and Dijkstra's Algorithm \cite{dijkstra1959note}. 
\updates{Trajectory planning has been studied extensively in the context of direct collocation \cite{schlossman2021open, kim2019flight, hargraves1987direct} , indirect (variational) methods \cite{betts1998survey, bonalli2019optimal}, sequential convex optimization \cite{wang2020improved, malyuta2021convex}, primitive-based planning \cite{bottasso2008path, goddard2021utilizing} and rapidly exploring random trees (RRTs) \cite{shi2020uav}.}
However, existing methods still suffer several drawbacks.  Solving optimization problems in real-time can still be computationally challenging.  In addition, sampling-based methods and discrete motion planners must be modified to account for kinodynamic constraints.  This can be done by performing closed loop forward simulations (CL-RRT) \cite{kuwata2009real} or using stereotyped behaviors such as motion primitives \cite{frazzoli2000robust}.  Motion primitives (MPs) are an attractive option because they are a type of high-level action that does not need to be specified at each time step of the horizon.  However, generating a motion primitive library can be time-consuming and non-intuitive.

Deep Reinforcement learning (RL) \cite{mnih2015human}  offers the potential to improve motion planning performance \cite{faust2018prm}.  One method of using Deep RL in motion planning is to generate end-to-end trajectories for systems with complex dynamics and constraints.  This leverages a key benefit of Deep RL, which is its ability to perform well for problems where a nonlinear programming formulation is not suitably defined.  Similarly, Deep RL can provide benefit when the underlying problem is stochastic \cite{bellemare2020autonomous}.  Deep RL has recently been applied in aerospace domains.  Deep RL is applied to spacecraft orbit guidance in \cite{lafarge2020guidance}, where rewards are carefully shaped to produce desirable behavior.  In \cite{hovell2020deep} Deep RL is applied to a spacecraft docking problem.  The recent works of \cite{waxenegger2021reinforcement, capobianco2021deep} apply Deep RL for control of rocket engines and Deep Learning for trajectory prediction using recurrent neural networks, respectively.

Despite much promise, Deep RL does not always provide effective end-to-end planning performance.  Specifically, RL can struggle when there are sparse rewards \cite{faust2018prm} or on long horizon tasks \cite{Qureshi2021Motion}.  In addition, achieving convergence often requires significant hyper-parameter tuning and reward shaping \cite{berner2019dota}.   As a result, many past works have used Deep RL to solve a smaller part of the planning problem.  For example, \cite{chiang2019RL-RRT} used Deep RL to estimate reachability, \cite{faust2018prm} used RL for local planning, and \cite{goddard2021utilizing}  used RL to learn motion primitives \updates{for use in graph-based planning}.  
  
High level actions are sub trajectories that have variable shape and duration.  These can then be transformed and concatenated to form end-to-end trajectories.  This differs from traditional guidance algorithms that update control inputs (in this case alpha and bank angle) continuously.  Using high-level actions can enable or improve different computational methods.  High-level actions \updates{have been examined in a range of contexts for separating the time scales of planning and control.}  For example, the maneuver automaton consists of trims and maneuvers that can be combined to create end-to-end kinodynamic plans \cite{frazzoli2000robust}.  Similarly, motions can be created using dynamic movement primitives (DMPs) \cite{ijspeert2002movement}.  These primitives are represented using dynamical systems theory and can be modulated and combined to achieve complex behaviors.  Each DMP is characterized by an intermediate goal parameter and shape parameters which decide how the intermediate goal is reached, \updates{along with a time-dilation parameter which specifies the duration of the movement.} DMPs have been applied to robotic flight in \cite{perk2006motion} where individual primitives are learned directly from training trajectory demonstrations.  \updates{DMPs have also been integrated with Deep RL in the context of robotic motion refinement \cite{kim2018learning}.  High-level actions in the form of sampled control action sequences have been shown to improve exploration in challenging aerial robotic environments \cite{dharmadhikari2020motion}.} 

\updates{In this paper we further examine how  high level actions can improve the performance of trajectory planners that are generated via RL.  We draw inspiration from DMPs and aforementioned works in high-level actions. We incorporate high-level actions as sub-trajectory splines of variable shape and duration, where both the shape and duration are designed by the RL agent at each segment along the horizon.  These sub-trajectory splines have been used in literature to represent different quantities, such as the control input sequence or desired system output trajectory. DMPs have previously been integrated into Deep RL \cite{saveriano2021dynamic}, and spline-based high-level actions have been known in the robotics community for some time \cite{peters2006policy}.  In this work we allow the planner (in our case, a Deep RL-trained planner) to choose the action duration at each segment along the planning horizon.  To our knowledge this is the first work which studies the RL agent actively choosing the duration for which the sub-trajectory is applied.}  
We show that our method is highly compatible with Deep RL, encouraging exploration within the environment, and we illustrate how only simple reward shaping is needed.  

\updates{An interesting comparison can be made between our work and Upside Down Reinforcement Learning (UDRL) \cite{srivastava2019training}. UDRL transforms the reinforcement learning problem into a supervised learning problem by separating the problem of data-generation and exploration from the problem of learning. In doing so, UDRL specifies a new input vector (containing desired horizon time) to a behavior function, which, in our case, would encode the time until the end of the current sub-trajectory. 	
The behavior function then outputs a high-level action which is passed through another function to obtain the final action, analogous to our use of a tracking controller (shown later in this work) to convert actions to control inputs.
Notably, in UDRL, the time horizons over which the algorithm learns to optimize are user-specified in advance, which distinguishes our work.}

Rapid planning for quasi-static or even static maps remains a challenging problem for systems with highly complex dynamics and constraints \cite{grant2016rapid}.  Traditional approaches for solving these types of problems in real-time often rely on receding horizon control where an optimization problem is performed on-line utilizing modeling simplifications \cite{kuwata2005robust}.  Alternatively, we utilize our Deep RL-based approach for trajectory planning where a state-feedback policy is first learned through off-line training.  Afterwards, this trained policy can be evaluated rapidly to produce system inputs in response to arbitrary state measurements.  \textcolor{black}{Our approach makes no assumptions on linearity or convexity \cite{malyuta2021convex}, and does not require an initial guess.}  Trajectory planning is demonstrated on two problems involving a shuttle reentry vehicle: the first problem is a free-space optimization problem, the second problem focuses on feasibility and incorporates a quasi-static map with obstacle regions.  On the first problem we demonstrate that our planner can achieve near-optimality compared to an optimal control solution.  On the second problem we demonstrate that our planner can rapidly and robustly produce feasible solutions under the quasi-static map.  Additionally, we present a new loss function term for policy gradient Deep RL.  This loss function term prevents the policy from producing actions which, on average, are outside action limits.  We describe how this method is analogous to the classical ``integrator windup'' problem in feedback control systems in which actuator limits are present. 
\begin{figure}
	\centering
	\begin{minipage}{.5\textwidth}
		\includegraphics[page=1, trim = 5mm 105mm 32mm 0mm, clip,width=.99\textwidth]{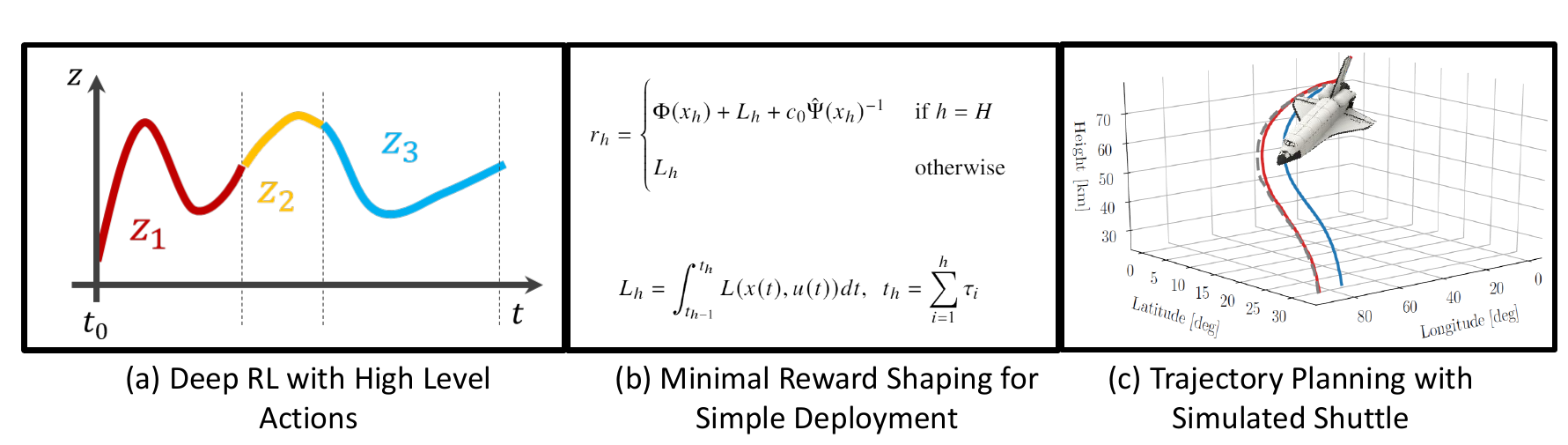}
	\end{minipage}
	\captionsetup{width=.49\textwidth}
	\caption{Overview of the approach.} 
	\label{fig:OverviewFig}
\end{figure}

\textbf{Contributions:}  The core contributions of this work are 1) \updates{incorporating a high-level action space (HLAS) for motion planning within Deep RL, where the agent chooses each action duration}, 2) illustration that this method promotes exploration and minimizes reward shaping, 3) the development of a loss function term to prevent ``action windup'' in policy gradient methods, and 4) illustrating performance path and training improvements with a shuttle re-entry example.  

This paper is organized as follows: A mathematical background is provided in Section II.  Section III describes the problem formulation where we describe our primary technique and frame the problem in the context of Deep RL.  We conclude the section with a short analysis describing the quality of solutions, and the influence the technique has on exploration in Deep RL problems.  Section IV describes a mechanism to prevent ``action windup'' in the context of Deep RL policy gradient methods.  Numerical experiments are performed on a shuttle reentry problem in Section V.  We present our conclusions in Section VI.

\section{Mathematical Background}
\subsection{Optimal Control}\label{section:Optimal Control}
Optimal control examines trajectory generation by considering the system dynamics and a reward function.  Optimal control seeks the time-series trajectory or feedback law that maximizes the reward function (or equivalently, minimizes a cost function).  The system dynamics are formulated as differential equations.  Consider a system with the following dynamics,
\begin{equation}\label{eq:dynamics}
	\dot x = f(x(t),u(t))
\end{equation}
where $x \in \mathbb{R}^n$ is the state and $u \in \mathbb{R}^m$ is the control.  With this system we associate the following performance reward function \cite{bertsekas1995dynamic}
\begin{align} \label{eq: J}
	J &= \Phi(x(t_H)) + \int_{t_0}^{t_H} L(x(t),u(t)) dt
\end{align}
where typically the initial time $t_0$ is fixed and horizon time $t_H$ is a variable to be optimized.  The instantaneous performance $L(x(t), u(t)) \in \mathbb{R}$ measures the performance of the system along the horizon, while $\Phi(x(t_H)) \in \mathbb{R}$ associates a final performance reward received at the end of horizon.  The goal is to design a control trajectory, $u(t):[t_0,t_H]\rightarrow \mathbb{R}^m$, which maximizes (\ref{eq: J}) subject to dynamics (\ref{eq:dynamics}) while simultaneously satisfying the following constraint conditions: 
\begin{enumerate}
	\item $u(t) \in [u_{min}, u_{max}] \subset \mathbb{R}^m ~\forall~ t\in[t_0,t_H]$
	\item $x(t)\in X \subset \mathbb{R}^n ~\forall~ t\in[t_0,t_H]$
	\item $g_j(x(t), u(t), t) \leq 0, ~j=1,\dots,J, ~\forall~ t\in[t_0,t_H]$
	\item $\Psi(x(t_H)) = 0, \Psi\in\mathbb{R}^q$
\end{enumerate}
The first two conditions ensures the control and state trajectories are \textit{admissible}, i.e., they do not violate any control or state constraints.  In this work we consider box form control constraints.  The third condition is a set of joint state-control path inequality constraints at each point along the horizon.  The last condition is a set of terminal constraints, which differs from the horizon reward function $\Phi(x(t_H))$.  In most cases the optimal control problem cannot be solved analytically.  Rather, so-called direct transcription methods \cite{kelly2017introduction} can be applied to solve the problem numerically, in which case the dynamics (\ref{eq:dynamics}) can be approximated with a numerical discretization of the form
\begin{equation} \label{eq:dynamics_discretetime}
	x_{k+1} = F_k(x_k,u_k)
\end{equation}
Here $F_k$ is some time-discretized approximation to the dynamics function $f$, and $x_k=x(k\Delta t)$, $u_k=u(k\Delta t)$ for some time step $\Delta t$ and $k\in \mathbb{Z}$. A distinct feature which many of these methods share is they are \textit{model-based}, whereby gradients of the dynamics and objective function are computed along trajectories based on analytical models.

\subsection{Deep Reinforcement Learning}\label{section: deep rl}
Rather than designing an optimal and admissible control trajectory, RL \cite{sutton2018reinforcement} seeks to design a feedback policy to maximize an associated performance index.  Unlike the model-based approach described in \ref{section:Optimal Control}, RL formulations can be \textit{model-free}, in which an \textit{agent} interacts directly with the \textit{environment}, periodically receiving rewards and state observations from the environment.  This is particularly valuable for environments that cannot be easily described by differential equations.  In this work, we refer to each step of the agent-environment interaction as an \textit{action step}, $h$, and distinguish it from the time step, $\Delta t$, associated with the discrete time approximation Eq. (\ref{eq:dynamics_discretetime}) of the dynamics.  It is through this direct interaction with the environment that the agent learns to adjust its policy to maximize the performance index.  In this work, we utilize RL to perform trajectory generation for nonlinear aerospace systems subject to complex constraints.

\subsubsection{Markov Decision Process}
In general, the RL problem is formalized by a discrete-time Markov Decision Process (MDP) consisting of the following components:
\begin{enumerate}
	\item The state observation space, $\mathcal{S}$, where each state $s\in \mathcal{S}\subset \mathbb{R}^{\dim(s)}$.  
	\item The set of feasible actions the environment can accept, called the \textit{action space}, $\mathcal{A}$.  We restrict our attention to continuous actions spaces, where $a\in\mathcal{A} \subset \mathbb{R}^{\dim(a)}$ is suitable for continuous control problems.  
	\item The state transition probability density function $p(s'|s,a)$ which describes the likelihood of the agent observing state $s'$ at action step $h+1$ given that the agent took action $a$ from state $s$ at action step $h$.
	\item The scalar function $r_h(s,a,s')$ is the expected reward received by the agent when taking action $a$ from state $s$ at action step $h$ and arriving at state $s'$ at action step $h+1$.  
\end{enumerate}
Associated with the MDP is the discounted return,
\begin{equation}\label{eq:discounted return}
	R_h = \sum_{k=h}^{H} \gamma^{k-h}r_{k}(s_k, a_k, s_{k+1})
\end{equation}  
Here $0 < \gamma \leq 1$ is the discount factor determining how much effect future rewards have on immediate decisions, and reduces the variance of the return when $\gamma < 1$ \cite{schulman2015high}.  We will show later that the discount factor brings additional benefits to our framework by encouraging exploration.  The agent-environment interactions may naturally separate into subsequences referred to as \textit{episodes}, in which case horizon length $H$ is a random variable which is determined by the agent reaching a terminal state.  This episodic case, in which there is a clear terminal objective, is our focus in this work. 

Actions are generated according to some policy $\pi$, which may be deterministic, $a = \pi(s)$, or stochastic, $a \sim \pi(a|s)$ in which case $\pi(a|s)$ represents the probability of choosing action $a$ when the state is $s$.  For any MDP there is an optimal policy which is deterministic \cite{littman1994markov}.  Nonetheless, defining the agent with a stochastic policy during training promotes exploration within the environment \cite{sutton2018reinforcement}.  The goal of an MDP is to design an optimal policy $\pi^* = \arg\max_\pi \E \left[ R_0 | \pi \right]$ which maximizes (\ref{eq:discounted return}) when actions are generated from the policy.

\subsubsection{Policy Gradient Methods}\label{section:pg_methods}
In this work we use \textit{policy gradient methods} \cite{peters2006policy}, where the policy is explicitly parameterized often as a deep feed-forward neural network, $\pi=\pi_\theta$. Here $\theta$ is a vector of adjustable network parameters.  A simple (and widely used) stochastic policy is constructed as a diagonal Gaussian distribution, where the actions are uncorrelated as all non-diagonal entries of the covariance matrix are zero
\begin{equation}\label{eq:GaussianPolicy}
	\begin{aligned}
		a &= \mu_\theta(s) + \sigma_\theta(s) \odot z \\
		z &\sim \mathcal{N}(0,I)
	\end{aligned}
\end{equation}
Here $\odot$ is element-wise multiplication, $z \in\mathbb{R}^{\dim(a)}$ is sampled from the standard normal distribution, while $\mu_\theta\in\mathbb{R}^{\dim(a)}$ and $\sigma_\theta \in\mathbb{R}^{\dim(a)}$ are the parameterized state-dependent action mean and standard deviation, respectively, so that $\pi_\theta(a|s)$ represents a diagonal Gaussian density. As an example, in the one-dimensional case we have $\pi_\theta(a|s)=\frac{1}{\sqrt{2\pi}\sigma_\theta(s)} \exp\left(-\frac{(a-\mu_\theta(s))^2}{2\sigma_\theta(s)^2}\right)$. \updates{The action mean $\mu_\theta(s)$ and standard deviation $\sigma_\theta(s)$ are deterministic functions dependent on parameters $\theta$ and state $s$.}  Policy gradient methods seek to maximize the expected return over actions and states induced from the policy and state transition distributions, respectively,
\begin{equation}
	\begin{aligned}
		L(\theta) = \E_{\substack{a\sim\pi_\theta(\cdot|s) \\ s'\sim p(\cdot|s,a)}} \left[R_1|\pi_\theta\right]
	\end{aligned}
\end{equation}
The policy parameters are adjusted through stochastic gradient ascent of the form $\theta_{k+1} = \theta_k + \alpha_{LR} \hat{g}(\theta)|_{\theta_k}$
where $\alpha_{LR}$ is the learning rate, and $\hat g(\theta) \approx \nabla_\theta L(\theta)$ is a sample average approximation of the policy gradient obtained by differentiating a surrogate objective function\footnote{For taking the derivative of $L^{PG}(\theta)$, $\log \pi_\theta(a|s)$ can be written in terms of $\mu_\theta(s)$ and $\sigma_\theta(s)$ and the dependence of these terms on $\theta$ is computed through automatic differentiation.} $L^{PG}(\theta)$.  This surrogate objective function is created over a finite batch of samples in an algorithm that alternates between sampling and optimization.  Several forms of the surrogate objective are available \cite{schulman2015high}, one of the simplest being
\begin{subequations}
	\begin{align}
		L^{PG}(\theta) 	
		&=  \frac{1}{N}\sum_{i=1}^{N}\left(\sum_{h=1}^H  R_h^{(i)} \log \pi_\theta(a_h^{(i)}|s_h^{(i)}) \right) \label{eq:L_pg}\\
		&= \hat \E_h\left[R_h \log \pi_\theta(a_h|s_h)\right] \label{eq:L_pg_expectation}
	\end{align}
\end{subequations}
where $s_h^{(i)}, a_h^{(i)}, R_h^{(i)}$ indicates states, actions and returns  from the $i^{th}$ trajectory sample at step $h$, and expectation $\hat\E_h[...]$ indicates the empirical average over a finite batch of samples $\{s_h,a_h,r_h,s_{h+1}\}$ collected by executing the policy within the environment. In section \ref{section:anti-windup} we formulate an inequality constrained optimization problem using a modified surrogate objective function.  
\subsubsection{Model-Free Learning}

\updates{Policy gradient methods are a type of model-free RL, meaning they make no attempt to explicitly learn the transition dynamics of the environment they are trained in.  Rather, these methods learn an optimal policy via direct interaction with the environment in a trial-and-error process. In many applications of Deep RL this environment is a simulated computational ``model'' used as a surrogate for the real system. Surrogate models are useful as performing millions of the required trial-and-error learning steps are prohibitive in terms of cost, safety and/or time in domains of interest such as stock market trading, flight, autonomous driving, medical applications, etc. 
These simulations are often parallelized and run significantly faster than real time such that years of real time training can be achieved risk free in a few hours or days. So-called sim2real techniques such as domain randomization, adversarial RL and transfer learning \cite{salvato2021crossing} have been developed to facilitate transfer to hardware when applicable. Though the environment is simulated based on known dynamics, only samples from these dynamics are used. In other words, the learning agent does not try to understand the computational model. In contrast, model-based RL explicitly learns a dynamics model as part of the learning process \cite{moerland2020model}. In comparison, model-free methods offer advantages due to their comparatively lower complexity \cite{lambert2020objective}. In this work simulated flight dynamics are the domain of interest. Ensuring alignment between those simulated dynamics and reality is left to future work.}

\section{Problem Formulation}\label{section: ProblemFormulation}
In this section we formulate trajectory generation as a Deep RL problem endowed with a high-level action space.  Trajectory generation is often formulated as an optimal control problem and then converted into a parameter optimization problem through direct transcription methods \cite{kelly2017introduction}.  This produces a problem formulation that is compatible with modern nonlinear programming techniques \cite{bertsekas2016nonlinear}. For many problems this method works well (see, e.g. \cite{schlossman2021open}), but long solution times and convergence stability can limit real-time application.  

In this work we develop an alternative method based on the notion of a continuously parameterized motion primitive \cite{ijspeert2002movement} and show the method is compatible with Deep RL.  We emphasize the novel use of a variable duration action, where each action-duration is uniquely chosen at every decision point along the horizon.  Our proposed method improves planning speed and stability by moving the computational burden off-line.  Difficult scenarios are solved off-line through randomized exploration of the state-space, and the best trajectory formulations are learned by the policy network.

\subsection{\updates{Spline Sub-Trajectories}}\label{section:HLAS}
\begin{figure}
	\centering
	\begin{minipage}{0.49\textwidth}
		\includegraphics [trim = 5mm 0mm 0mm 0mm, clip,width=.99\textwidth]{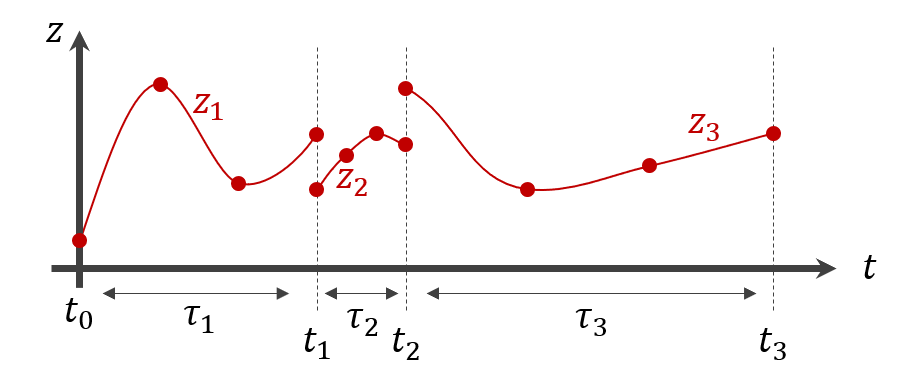}
	\end{minipage}
	\captionsetup{width=.49\textwidth}
	\caption{Example piecewise polynomial construction ($p=3$) for $z(t)$ for a horizon with three segments, $H$=3.  $t_{h} = \sum_{i=1}^{h}\tau_i$.}
	\label{fig:z visual}
\end{figure}

Consider a system with dynamics as described by (\ref{eq:dynamics}) and a subset of the state $y=Cx\in\mathbb{R}^\ell$ corresponding to the system output.  We define actions as sub-trajectories of variable duration.  Rather than requiring an action decision to be made at every time step, a constrained input sequence executes over multiple time steps and action decisions are only required between sub-trajectories.  In this work, the sub-trajectory at action-step $h$, denoted as $z_h$, can represent two quantities:
\begin{enumerate}
	\item A control input function, $u(t)=z_h(t)$
	\item A desired \updates{output function}, $\dot y^{des}(t)=z_h(t)$.  
\end{enumerate}
For the second case it is assumed a controller can be designed so the output follows the \updates{desired output} function.  A tracking controller for this task is discussed in Section \ref{section:NumericalExp_TrackingController}.
In each case $t_{h-1} \leq t \leq t_{h}, ~~t_{h} = \sum_{i=1}^{h}\tau_i$, and $\tau_h$ is the duration of action-step $h$ for $h=1,\dots, H$.  Here $H$ is the terminal step, a random variable identical to that described in Section \ref{section: deep rl}.  Each sub-trajectory is a time-varying function parameterized as a polynomial of degree $p \geq 0$ 
	\begin{align}\label{eq:z}
		z(t) &= z_h\left(\frac{t-t_{h-1}}{\tau_h}\right), ~~~
		z_h(t') = \sum_{k=0}^{p} c_{h,k+1} t'^k
	\end{align}
for $h=1,\dots,H$, $t_0=0$, and $0\leq t' \leq 1$.  The polynomial degree, $p \geq 0$, is a design choice made before the problem is solved.  $p+1$ nodes are required to uniquely specify a $p^{th}$ order polynomial.  Nodes $n_{h,1}, \dots, n_{h,p+1}$ are evenly placed along each sub-horizon $\tau_h$ with node $n_{h,1}$ placed at the start of the sub-horizon and $n_{h,p+1}$ placed at the end of the sub-horizon.  \updates{It is noted that each polynomial can be evaluated anywhere along $[t_{h-1}, t_h]$ during implementation, i.e., the evaluation points are not limited to the evenly spaced nodes.} The $p+1$ coefficients $c_{h,1},\dots, c_{h,p+1}$ for each step $h$ along the horizon are designed through an appropriate curve fitting interpolation. 
Decision variables are: 
\begin{enumerate}
	\item the $p+1$ nodes $n_{h,1},\dots, n_{h,p+1}$ for each step $h=1,\dots,H$ along the horizon, and
	\item the \textit{action duration}, $\tau_h$, over which the polynomial profile is to be applied
\end{enumerate}
Figure \ref{fig:z visual} visually describes the formation of $z(t)$ with variable length segments $z_h$.  
As evident in Fig. \ref{fig:z visual} discontinuities in $z$ are permitted in the most general form.  However, during implementation an additional constraint of the form $z_{h+1}(0)=z_{h}(1)$ can be added to ensure continuity across sub-trajectories.  

In the case where $z = \dot y^{des}$, for a $p^{th}$ order function $z(t)$ we are effectively defining piecewise polynomial desired output profiles, $y^{des}(t)$, of degree $p+1$,
\begin{equation}\label{eq:y^des}
	\begin{aligned}
		y_h^{des}(t) 
		&= y_{h-1}^{des}(t_{h-1}) + \int_{0}^{(t-t_{h-1}) / \tau_h} z_h(s) ds \\
		&= y_{h-1}^{des}(t_{h-1}) + \sum_{k=0}^{p} \frac{1}{(k+1)} c_{h,k+1} \left(\frac{t-t_{h-1}}{\tau_h} \right)^{k+1} 
	\end{aligned}
\end{equation}
where $t_{h-1} \leq t \leq t_h ~~,~~ t_{h} = \sum_{i=1}^{h}\tau_i$.
\subsection{Formulation as a Deep RL Problem} \label{section: HLAS Deep RL Formulation}
\subsubsection{Rewards}
The reward signal of the discounted return (\ref{eq:discounted return}) is defined as
\begin{equation}\label{eq:HLAS Deep RL reward}
	r_h = 
	\begin{cases}
		\Phi(x_h) + L_h + C_0\hat\Psi(x_h) &\text{ if $h=H$} \\
		L_h &\text{ otherwise}
	\end{cases}	
\end{equation}
where
\begin{equation} \label{eq:HLAS Deep RL reward performance index}
	L_h = \int_{t_{h-1}}^{t_h} L(x(t),u(t))dt,~~t_h=\sum_{i=1}^{h}\tau_i
\end{equation}
is the performance index accumulated along action step $h$, and $L(x,u), \Phi(x_h)$ are from (\ref{eq: J}).  We have relaxed the terminal constraint function $\Psi=0$ from section \ref{section:Optimal Control} as a \updates{reward of the form $C_0\hat\Psi(x_h)$, where $C_0$ is a tunable constant and $\hat\Psi(x_h)$ is an indication of convergence to $\Psi(x_h)=0$.  A larger value of $\hat\Psi(x_h)$ indicates better convergence to $\Psi(x_h)=0$ and we therefore seek to maximize $\hat\Psi(x_h)$.}
\subsubsection{Action Space}\label{section: ProblemFormulation_actionspace}
The action space of the RL agent is defined as $\mathcal{A}=(\tau, n_{1},\dots,n_{p+1})$ as described in section \ref{section:HLAS}.  At each action step the agent chooses the time duration $\tau$ of the step and the location of the $p+1$ evenly spaced node points $n_i$ as shown in Fig \ref{fig:z visual}.  Lower and upper limits are set on the action duration: $\tau_{min} \leq \tau \leq \tau_{max}$.  Additionally, lower and upper limits are placed on each node point so that each $z_h$ remains bounded:  $z_{min} \leq n_{i} \leq z_{max}, ~i={1,\dots,p+1}$.
\subsubsection{Constraints}
The box form control input constraints, $u\in U$, are enforced directly in the environment through clamping. The control inputs are computed based on the actions and the tracking controller.  For the state constraint, $x\in X$, and path constraints, $g_j(x,u)\leq 0, j=1,\dots,J$, we employ a simple strategy in which the episode is ended and the agent receives no further reward if the agent violates any of these constraints.  Over the course of learning, the agent learns to avoid regions of infeasibility.

\subsubsection{HLAS Diagram}
The figure below shows the HLAS Deep RL agent-environment interaction loop. At every action-step $h$ the agent generates an action $a_h=(\tau_h, \{n_{h,i}\})$ which specifies the profile of $z_h(t)$ along $t_{h-1} \leq t \leq t_h$ where $t_h=\sum_{i=1}^{h}\tau_i$.  The number of timesteps $dt$ between action-steps $h$ is determined by $\tau_h$.  
The tracking controller converts $y^{des}$ or $\dot y^{des}$ into a control input $u$ depending on the definition of $z$ (when $z_h(t)=u(t)$ the tracking controller is the identity function).  
The environment (which is comprised of the $z_h$ profile generator, the tracking controller, the dynamics, and the reward definition) provides the agent with an updated reward and state observation every action-step.  

\begin{figure}
	\centering
	\begin{minipage}{0.49\textwidth}
		\includegraphics [trim = 0mm 0mm 0mm 0mm, clip,width=.99\textwidth]{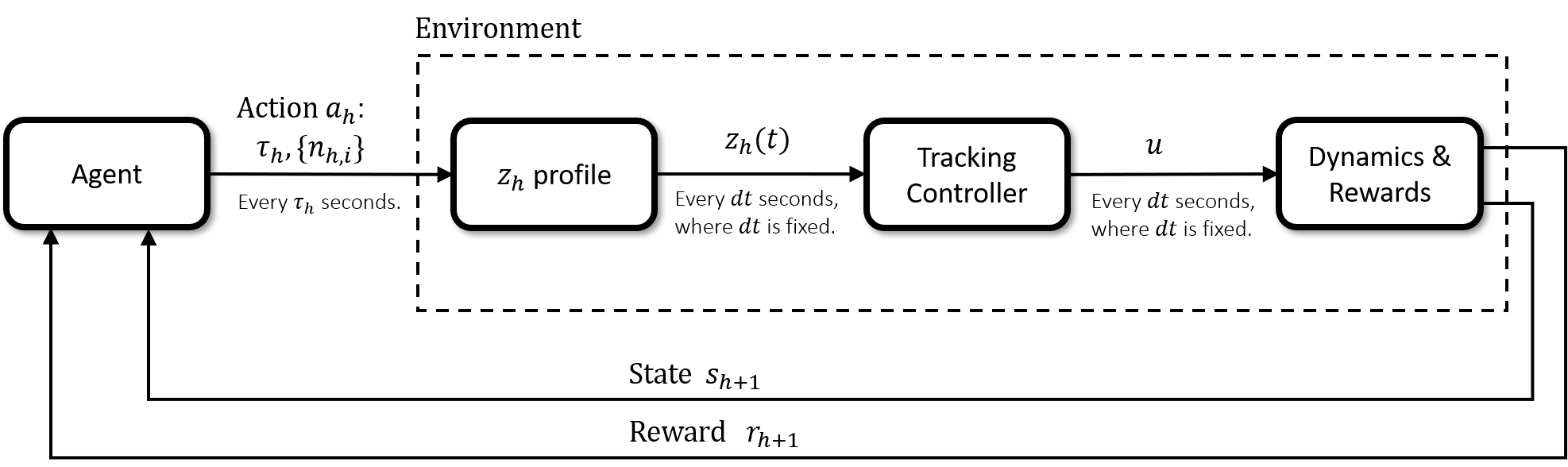}
	\end{minipage}
	\captionsetup{width=.49\textwidth}
	\caption{Implementation of the HLAS Deep RL formulation.} 
	\label{fig:HLAS_diagram}
\end{figure}

\subsection{Analysis}\label{section:Analysis}
\subsubsection{Accuracy of Approximation}
The method described in Section \ref{section:HLAS} is a way to parameterize a trajectory generation problem for Deep RL.  It can also be viewed as a way to approximate a solution to the optimal control problem in Section \ref{section:Optimal Control}.  In Section \ref{section:DebrisAvoidance} we will use the methods described in Section III to perform trajectory planning on a problem which is challenging for traditional optimal control solvers. The following shows that the best approximation error grows at a bounded linear rate along the horizon in the case where $z=\dot y^{des}$. 

\textbf{Lemma} (Accuracy of HLAS approximation, for $z=\dot y^{des}$).
Define $x^*:[t_0,t_H]\rightarrow\mathbb{R}^n$, $u^*:[t_0,t_H]\rightarrow\mathbb{R}^m$ as a solution to (\ref{eq:dynamics}) which is optimal with respect to (\ref{eq: J}).  Define $\dot {\hat x}(t)=z^*(t)$, where $z^*(t)$ is a piecewise polynomial as described in (\ref{eq:z}) over $t\in [t_0,t_H]$ and best approximates $f(x^*(t),u^*(t))$ over $t\in [t_0,t_H]$ in some sense.  Assume $\hat x(t_0) = x^*(t_0)$ is a specified initial condition.  Furthermore assume $\Vert z^*(t) - f(x^*(t), u^*(t))\Vert_\infty \leq m_1 ~\forall t \in [t_0,t_H]$, for some $m_1>0$.  Then $\Vert \hat x(t) - x^*(t) \Vert_\infty \leq m_1[t-t_0] \forall t \in[t_0,t_H]$.

\textit{Proof}.  $\hat x(t) - x^*(t) = \int_{t_0}^{t} z^*(\tau) d\tau + \hat x(t_0) - \int_{t_0}^{t} f(x^*(\tau),u^*(\tau)) d\tau - x^*(t_0) = \int_{t_0}^{t} z^*(\tau) - f(x^*(\tau),u^*(\tau)) d\tau $.  We then have $\Vert \hat x(t) - x^*(t) \Vert_\infty = \int_{t_0}^{t} \Vert z^*(\tau) - f(x^*(\tau),u^*(\tau)) \Vert_\infty d\tau \leq \int_{t_0}^{t} m_1 d\tau = m_1 [t-t_0]$.  

\textbf{Remark} (Bounds on $m_1$).  Bound each component of $f(x^*(t),u^*(t))$ as $L_i(t_h) = \inf\limits_{t_h \leq t \leq t_h+\tau_h} f_i(x^*(t),u^*(t))$, 
$U_i(t_h) =  \sup\limits_{t_h \leq t \leq t_h+\tau_h} f_i(x^*(t),u^*(t))$,
where $\tau_h$ is the action duration as described in Section \ref{section:HLAS}.  Set $z_i(t) = \frac{U_i(t_h)-L_i(t_h)}{2}$ as a suboptimal approximation to $f_i(x^*(t),u^*(t))$ along $t \in [t_h,t_h+\tau_h]$.  We then have $\vert z_i(t)-f_i^*(t) \vert  \leq \frac{\Delta_i}{2}$, where $\Delta_i=\sup_{h} (U_i(t_h) - L_i(t_h))$.  We can therefore take $m_1=\sup_i\frac{\Delta_i}{2}$.

The bound $\Delta_i$ will generally decrease as $\tau_h$ is decreased by the definitions of $L_i(t_h)$ and $U_i(t_h)$.  Furthermore, by the Weierstrass Approximation Theorem, $f_i(x^*, u^*)$ can be approximated to arbitrary accuracy by a polynomial of sufficient degree.  Therefore, $\Delta_i$ will generally decrease as the polynomial degree $p$, as described in Section \ref{section:HLAS}, is increased.  

\subsubsection{Effect on Environment Exploration}\label{section:episodic reward}
The variable action duration $\tau_h$ has interesting implications in terms of environment exploration during the RL training process.  We now examine the effect of discounting on the action duration for a specific reward structure, and show longer action durations are preferred. Longer action durations induce larger movements through the state space between gradient updates, promoting environment exploration and ultimately producing improved policies.  This effect is empirically demonstrated in Section \ref{section:shuttle_ablation_studies}.

\textbf{Definition} (Strictly Episodic Reward).  A reward that is guaranteed to occur every episode, but only at a terminal state: $r_h=0 ~\forall~ h \neq H$, $r_H > 0$, where $H$ is the horizon length described in Section \ref{section: deep rl}.

When positive and strictly episodic rewards are used in conjunction with the HLAS Deep RL formulation, a discount factor $0 < \gamma < 1$ encourages the agent to favor larger action durations $\tau_h$.  The discounted episode return is $R=\sum_{h=1}^{H} \gamma^{h-1}r_h = \gamma^{H-1} r_H$, where the term $\gamma^{H-1}$ is the amount by which the episodic reward $r_H$ is discounted \updates{($r_h=0~ \forall h\neq H$ assuming strictly episodic rewards)}.  Consider two horizon lengths $H_1, H_2$ where $H_1<H_2$.  If $r_{H_1} = r_{H_2}$, horizon length $H_1$ will be preferred by the agent as this produces a larger discounted episode return.  As a result, the agent will attempt to keep $H$ small by choosing larger action durations $\tau_h$.

\subsection{Deep RL Training}\label{section:RLTrainingSetup}
\begin{figure}[h!]
	\centering
	\begin{minipage}{0.40\textwidth}
		\includegraphics [trim = 0mm 0mm 0mm 0mm, clip,width=.99\textwidth]{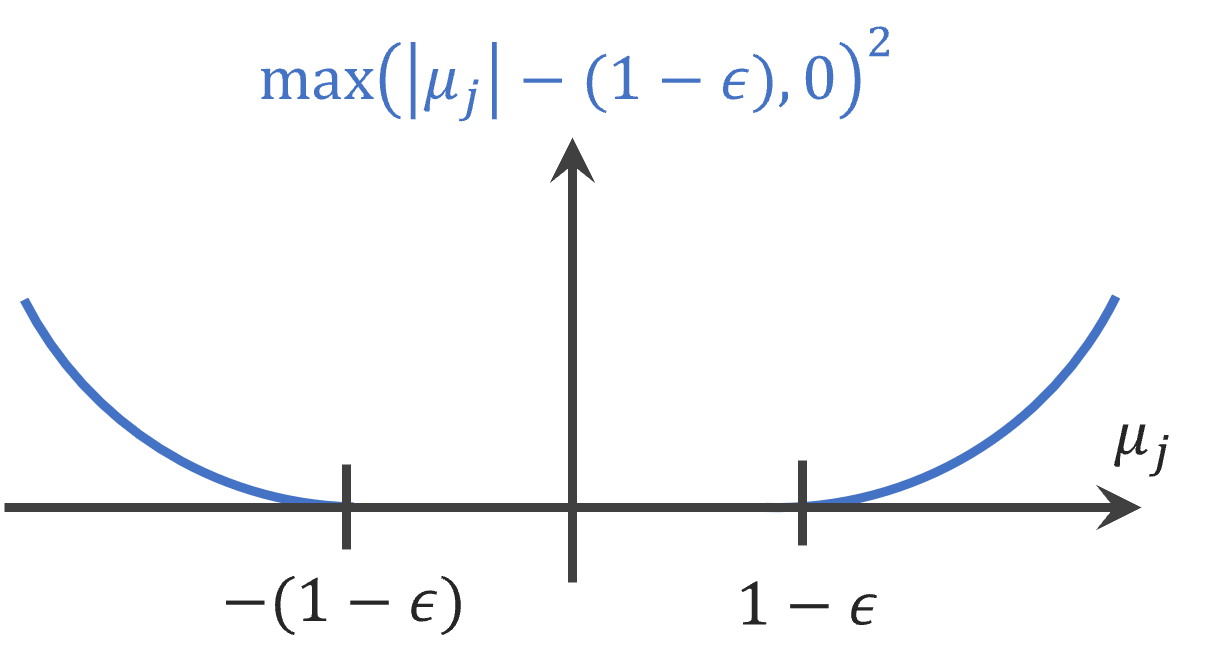}
	\end{minipage}
	\captionsetup{width=.49\textwidth}
	\caption{The inner max-squared term from (\ref{eq:AntiWindupLossTerm}).}
	\label{fig:loss_function_penalty_term}
\end{figure}

\updates{The goal of Deep RL training is to tune policy parameters $\theta$ so the average episode return is maximized. For this task} a range of algorithms can be used including off-policy algorithms such as Soft Actor Critic (SAC) \cite{haarnoja2018soft}, Deep Deterministic Policy Gradient (DDPG) \cite{silver2014deterministic}, 
and on-policy algorithms such as 
%
%
Asynchronous Advantage Actor Critic (A3C) \cite{mnih2016asynchronous} and Proximal Policy Optimization (PPO) \cite{schulman2017proximal}.  While off-policy methods tend to be more sample efficient (since they can reuse old data through the use of a replay buffer), on-policy methods directly optimize the objective and favor stability over sample efficiency \cite{SpinningUp2018,tang2020discretizing}.  
In particular, PPO is an on-policy algorithm which restricts the size of policy update via a KL divergence penalty or clipping mechanism.  PPO is one of the most popular Deep RL methods, achieving state-of-the-art performance across a wide range of challenging tasks \cite{wang2019trust,schulman2017proximal}.  In this work we use PPO for all Deep RL training.  Our implementation of PPO uses the following surrogate objective function, 
\begin{equation}\label{eq:PPOSurrogateObj}
	\begin{aligned}
		L^{PPO}(\theta) &= \hat\E_{h}\big[ -L_h^{\text{clip}}(\theta) + C_1 L_h^{VF}(\theta) - C_2 S[\pi_\theta](s_h) + \\
		&~~~~~~~~~~~~~ L_{\mu_\theta}(s_h) \big] 		
	\end{aligned}
\end{equation}
Here $\hat \E_h[f(s_h,a_h)]$ indicates the sample average approximation to 
$\E_{s\sim \rho^{\pi_\theta}(\cdot),a\sim\pi_\theta(\cdot|s)} \left[f(s,a)\right]$
where $\rho^{\pi_\theta}(s)$ is the distribution of states induced by the policy \cite{sutton1999policy}.  The terms $L_h^{\text{clip}}(\theta)$, $L_h^{VF}(\theta)$ and $S[\pi_\theta](s_h)$ are the standard PPO-clip, value function and entropy exploration bonus terms, respectively, taken directly from \cite{schulman2017proximal}. The entropy term is included as it has been shown to improve exploration by discouraging premature convergence to suboptimal policies \cite{mnih2016asynchronous}. Coefficients $C_1$ and $C_2$ are hyperparameters chosen heuristically.
The last term in the objective is a contribution in this work for bounding the learned action distribution,
\begin{equation}\label{eq:AntiWindupLossTerm}
	L_{\mu_\theta}(s_h) = \hat\E_{h}\left[\sum_{j=1}^{\dim(a)}c_j \max\left(\left|\mu_{j,\theta}(s_h)\right|-(1-\epsilon), 0\right)^2 \right]
\end{equation}
\updates{Here, $\mu_j$ represents the mean of the $j^{th}$ action dimension, and $\mu_{j,\theta}(s_h)$ is the network output of the $j^{th}$ action mean given state $s_h$, as described in section \ref{section:pg_methods}. This loss term and a method for automatically adjusting coefficients $c_j$ will be derived in Section \ref{section:anti-windup}. The form of the inner squared term is shown in Fig. \ref{fig:loss_function_penalty_term}.}

The PPO algorithm alternates between data collection and policy optimization.  During data collection the policy $\pi_\theta$ is executed on multiple instances of the environment in parallel \cite{mnih2016asynchronous, schulman2017proximal}, performing a fixed number of action-steps on each parallel environment.  During optimization stochastic gradient descent is performed by cycling through minibatches of the collected data to obtain sampled estimates of $\nabla_\theta L^{PPO}(\theta)$.  The final output of DeepRL training is a fully trained policy network $\pi_\theta(a|s)$.  In this work we use the Stable Baselines3 codebase \cite{stable-baselines3} for implementation of the PPO algorithm. After training is complete, the trained policy can be used to produce actions $a$ from any state $s$.  Referring to Eq. (\ref{eq:GaussianPolicy}), rather than stochasticaly producing actions as $a = \mu_\theta(s) + \sigma_\theta(s) \odot z$ (where $z \sim \mathcal{N}(0,I)$), we instead produce actions deterministicly from the action-mean portion of the policy network only, $a=\mu_\theta(s)$.

\section{Preventing Action Windup in Stochastic Policy Gradients}
Almost all physical systems have action limits due to hardware limitations.  These can be limits on control surface deflections or actuator force magnitudes.  When an agent exceeds these bounds during learning, problems can arise.  In this section we introduce a novel method for bounding the learned action distribution in policy gradient methods.  We focus on settings where the action the environment can accept is limited with box type bounds, producing a ``clipped'' action.  We demonstrate the effectiveness of this method in Section \ref{section:TraversedLatitude}.

The issues associated with bounded actions in Deep RL policy gradients has been explored previously.  One approach is to use the clipped action, referred to here as $\hat a_h$, directly in the policy gradient estimate in place of the unclipped action $a_h$.  Unfortunately this has been shown to introduce significant bias into the policy gradient estimation process \cite{chou2017improving} by effectively corrupting the empirical average (\ref{eq:L_pg}).  The clipped action policy gradient method \cite{fujita2018clipped} attempts to alleviate this bias issue by replacing the gradient with calculations based on the cumulative distribution function (CDF).  However, computing the CDF can be non-trivial (such as with Gaussian distributions where no closed form exists).  The work of \cite{haarnoja2018soft} employs the use of a $\tanh$ \textit{squashing function} to transform an unbounded Gaussian distribution to a bounded distribution as $\hat a_h = \tanh(a_h)$, where $a_h$ is the action produced by the (Gaussian) policy (\ref{eq:GaussianPolicy}).  However, the $\tanh$ transformation does not solve the ``windup issue,'' as the input to the $\tanh$ function can grow without bound with no further effect on the output.  The work of \cite{chou2017improving} investigates the applicability of the Beta distribution, which has bounded support, to represent the stochastic policy.  One drawback of the Beta distribution is that the policy must be nearly deterministic for near-limit action values to be chosen with high probability \cite{fujita2018clipped}.

\subsection{The Windup Problem}
\begin{figure}
	\centering
	\begin{minipage}{0.49\textwidth}
		\includegraphics [trim = 0mm 0mm 0mm 0mm, clip,width=.99\textwidth]{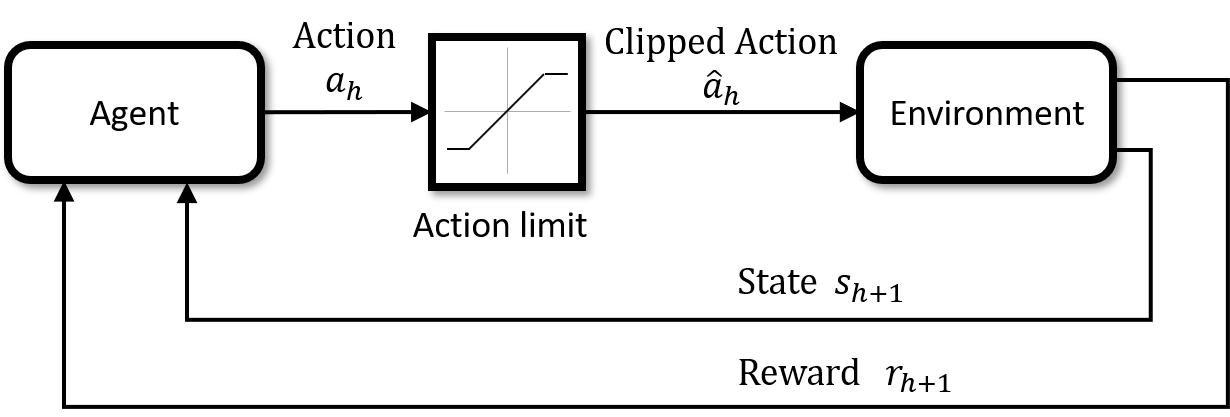}
	\end{minipage}
	\captionsetup{width=.49\textwidth}
	\caption{The agent–environment interaction in reinforcement learning, with action limits explicitly shown separated from the environment.}
	\label{fig:agent-env-interaction}
\end{figure}

Consider the typical agent-environment reinforcement learning training process \cite{sutton2018reinforcement}, shown in Fig. \ref{fig:agent-env-interaction} for box-type action constraints.
Here we have separated the action limits from the environment for clarity in our explanation.  This interaction can be viewed as a typical closed-loop feedback control process \cite{aastrom2021feedback} where the agent serves the purpose of a typical controller and the environment consists of the system being controlled, including all measurements.  The presence of the action limit in the forward path of the loop can be problematic.  For systems which operate over a wide range of conditions, the action delivered by the agent may very well reach a physical bound represented here by the action limit.  When this happens,  the feedback loop is broken and the process runs in open-loop, independent of the feedback from the environment \cite{aastrom2021feedback}.  For feedback controllers which possess an integrator process (i.e. for controllers which have a form of memory) the consequence can be long and undesirable transient behavior, which is sometimes destabilizing.  In the feedback control literature this well-studied phenomenon is referred to as \textit{integrator windup}.  \textcolor{black}{In the context of Deep RL, windup can occur in the action mean $\mu_\theta$ as it approaches the action limit.  Actions are sampled from the Gaussian policy, $a_h \sim \pi_\theta(\cdot|s_h)$, and all samples outside the limit will have the same effect on the environment as actions at the limit.  If a favorable reward $r_{h+1}$ is consistently received at the action limit at some point during training, then the aggregate effect of a policy gradient method will push the action mean $\mu_\theta$ towards and even well beyond the limit.  This can be problematic if the probability of sampling an action within the action limit becomes sufficiently low (due to the drifting $\mu_\theta$), since at this point the agent has effectively stopped exploring the environment.}  

In feedback control literature, a well-established method of \textit{anti-windup} involves making an adjustment to the input of the feedback controller based on the difference between the pre-limited and post-limited actions \cite{aastrom2021feedback}.  We now introduce an analogous method for policy gradient methods in Deep RL.
\subsection{Inequality Constrained Policy Optimization for Anti-Windup} \label{section:anti-windup}
\updates{In this section we derive the loss term shown in (\ref{eq:AntiWindupLossTerm}).}
\subsubsection{KKT Conditions}
\updates{Consider the following inequality constrained optimization problem, with $L^{PG}(\theta)$ as in (\ref{eq:L_pg})}
\begin{subequations}\label{eq:ConstrainedSurrogateOpt}
	\begin{align}
		\min_\theta&~~~ -L^{PG}(\theta) \\
		\text{subject to}&~~~ \hat\E_{h}\left[\mu_{j,\theta}^2(s_h)-1\right] \leq 0 \label{eq:Inequality}
	\end{align}
\end{subequations}
for $j=1,\dots,\dim(a)$, where $\hat\E_h[\dots]$ indicates the sample average with respect to quantities indexed by $h$, and $\mu_{j,\theta}(s_h)$ denotes the mean of action $j$ \updates{which is dependent on state $s_h$}.  The inequality constrained optimization (\ref{eq:ConstrainedSurrogateOpt}) seeks to optimize the standard surrogate objective function $L^{PG}(\theta)$ while maintaining the mean of each action dimension bounded between $[-1,1]$.  We construct the following Lagrangian and associated Karush–Kuhn–Tucker (KKT) \cite{bertsekas2016nonlinear} first order necessary conditions,
\begin{subequations}
	\begin{align}
		\mathcal{L}(\theta) &= -L^{PG}(\theta) + \sum_{j}c_j  \hat\E_{h}\left[\mu_{j,\theta}^2(s_h)-1\right] \label{eq:KKT Lagrangian} \\
		0 &= \left.\nabla_\theta \mathcal{L}(\theta)\right|_{\theta^*,c_j^*}  \label{eq:KKT stationary point} \\	
		0 &= c_j^* \hat\E_{h}\left[\mu_{j,\theta^*}^2(s_h)-1\right] \label{eq:KKT complimentary} \\
		0 & \leq c_j^* \label{eq:KKT positive cj}
	\end{align}
\end{subequations}
for $j=1,\dots,\dim(a)$. We next devise a scheme to approximately solve for the conditions (\ref{eq:KKT stationary point})-(\ref{eq:KKT positive cj}) under stochastic gradient ascent optimization. 

\subsubsection{Approximately Solving for the KKT Conditions} \label{section:anti-windup_approxKKT}
We adopt a penalty function approach from constrained optimization \cite{bertsekas2016nonlinear} in which case the KKT conditions are approximately satisfied under stochastic gradient ascent optimization.  The main idea is the inequality constraint (\ref{eq:Inequality}) is replaced with a positive penalty function which resembles the inequality, and penalty coefficients are adapted sequentially.  We modify the Lagrangian, introducing a penalty function of the form $P(\mu)=\max\left(\left|\mu\right|-1, 0\right)^2$ \updates{which is illustrated in Fig. \ref{fig:loss_function_penalty_term}}, 
\begin{equation}\label{eq:KKT_Approx Lagrangian}
	\begin{aligned}
	\mathcal{L}(\theta) &= -L^{PG}(\theta) + L_{\mu_\theta}(s_h) \\
	 L_{\mu_\theta}(s_h) &= \sum_{j}c_j  \hat\E_{h}\left[ \max\left(\left|\mu_{j,\theta}(s_h)\right|-1, 0\right)^2 \right]
	\end{aligned}
\end{equation}
This penalty function has several desirable properties: it has continuous derivatives with respect to $\mu$ at $\mu=+/-1$, $P(\mu)\geq 0$ for all $\mu$, and $P(\mu)=0$ if and only if $\mu\in[-1,1]$.
\updates{These properties of $P(\mu)$ increase the sensitivity of the empirical average $\hat\E_{h}\left[P(\mu_{j,\theta}(s_h))\right]$ to samples $\mu_{j,\theta}(s_h)$ whose magnitude are larger than 1. This is particularly useful in situations where most samples $\mu_{j,\theta}(s_h)$ are well within $[-1,1]$.}
Condition (\ref{eq:KKT complimentary}) is the so-called complimentary slackness condition, stating that $c_j^*$ must be zero whenever the inequality in (\ref{eq:Inequality}) is inactive.  We relax (\ref{eq:KKT complimentary}) as
\begin{equation}\label{eq: KKT Approx complimentary}
	\begin{aligned}	
	c_j^* \hat\E_{h}\left[ \max\left(\left|\mu_{j,\theta^*}(s_h)\right|-1, 0\right)^2 \right] = 0 
\end{aligned}
\end{equation}
for $j=1,\dots,\dim(a)$, noting that any ($\theta^*, c_j^*$) which satisfy (\ref{eq:KKT complimentary}) automatically satisfy (\ref{eq: KKT Approx complimentary}).  The conditions specified in (\ref{eq:KKT stationary point}) and (\ref{eq:KKT positive cj}) place further restrictions on the choice of $\theta^*$ and $c_j^*$.  Specifically (\ref{eq:KKT stationary point}) states that the optimal values for $\theta^*$ and $c_j^*$ must be concurrently chosen to produce a stationary point in the surrogate objective.  For this we employ two processes.  First, standard stochastic gradient ascent is applied to the surrogate objective (\ref{eq:KKT_Approx Lagrangian}) to adjust parameters $\theta$.  Second, at each gradient step, the coefficients $c_j$ are adapted according to a method similar to that developed in \cite{schulman2017proximal}\footnote{The method developed in \cite{schulman2017proximal} is for adapting a coefficient in KL divergence targeting.}, \updates{shown in Algorithm \ref{algorithm:cj_adaptation}.  On line 11, the gradient can be computed automatically using industry standard \textit{autograd} methods \cite{NEURIPS2019_9015}.  Constants $\alpha$ and $\beta$ are set to 1.5 and 2, respectively.  These values were empirically determined, providing sufficiently fast adaptation of the penalty coefficients without inducing oscillation.  
}  
\begin{algorithm}[] 
	\SetAlgoLined
	\KwIn{
		\\~~~Constants $\alpha_{LR}, d_{tar}$, $\alpha$, $\beta$ 
		\\~~~Batch data $\{s_h, a_h, r_{h}, s_{h+1}\}$
		\\~~~Initial values for $c_1,...,c_{\dim(a)}, \theta$} 
	\For{j = 1,2,...,$\dim(a)$}{
		$d_j = \hat\E_{h}\left[ \max\left(\left|\mu_{j,\theta}(s_h)\right|-1, 0\right)^2 \right]$ \\
		\If{$d_j < d_{tar} / \alpha$}{$c_j \leftarrow c_j / \beta$}
		\If{$d_j > d_{tar} * \alpha$}{$c_j \leftarrow c_j * \beta$}
	}
	{$\mathcal{L(\theta)} = -L^{PG}(\theta) + \sum_j c_j d_j$} ~\#$L^{PG}(\theta)$ from (\ref{eq:L_pg_expectation})\\
	{$\hat g = \nabla \mathcal{L}(\theta)$ }	\\
	{$\theta \leftarrow \theta - \alpha_{LR} \hat{g}$}
	\caption{Policy gradient step with penalty adaptation}\label{algorithm:cj_adaptation}
\end{algorithm}
This adaptation process automatically satisfies (\ref{eq:KKT positive cj}), assuming each $c_j$ is initialized with some positive value.  Examining our composite surrogate objective (\ref{eq:KKT_Approx Lagrangian}), the second term places external pressure on the network optimization process to find values $\theta$ which, on average, keep the action means bounded within the feasible range of the action space.

\subsubsection{Implementation Note}
In actual implementation we slightly modify the summation term in (\ref{eq:KKT_Approx Lagrangian}) as 
$c  \hat\E_{h}\left[\sum_{j} \max\left(\left|\mu_{j,\theta}(s_h)\right|-(1-\epsilon), 0\right)^2 \right]$.
For one, we have made the simplification of summing over all action dimensions inside the empirical average.  As a result, we only need to adapt one coefficient\footnote{For some applications there may be benefit in adapting all coefficients $c_j$, $j=1,\dots,\dim(a)$, but in this work we found the simplification to a single coefficient acting upon a summation of terms to be sufficient.}, $c$, using $d=\hat\E_{h}\left[\sum_{j} \max\left(\left|\mu_{j,\theta}(s_h)\right|-(1-\epsilon), 0\right)^2 \right]$ in place of $d_j$ in the update scheme above.  Second, we use $(1-\epsilon)$ with $\epsilon\approx 0.1$ inside the max function and set $d_\text{tar}=\epsilon^2$.  This allows us to target a positive value of $d_\text{tar}$ in the update scheme above while seeking $\hat\E_h\left[\sum_{j=1}^{\dim(a)}\left|\mu_{j,\theta}(s_h)\right|\right] \leq 1$. 

\section{Numerical Experiments with the Space Shuttle Reentry Problem}\label{section:NumericalExp}
In this section we evaluate the performance of our approach using the well-studied space shuttle reentry problem \cite{betts2010practical}.  We benchmark our approach against a state-of-the-art nonlinear programming solution, and a straightforward implementation of Deep RL in which action decisions are made by the RL agent every simulation time step.  A trajectory planning problem is also studied where the reentry must be planned over a wide range of initial conditions, avoiding a quasi-static group of obstacles within the map.

Deep RL training is performed using the PPO algorithm as described in Section \ref{section:RLTrainingSetup}.  We use six\footnote{Training was performed on a machine with a six core CPU.} parallel instances of the environment with 4096 action-steps per environment, and minibatch sizes of 128 action-steps during the optimization phase unless specified otherwise.  We use a network with two shared hidden layers of size 256, followed by an additional hidden layer of size 256 which is evenly split between the policy and value function approximation \cite{stable-baselines3, mnih2016asynchronous}.  The rectified linear unit (ReLU) activation function was used in each layer, various hyperparameters used throughout training are listed in the following sections.  The shuttle dynamics Eq. (\ref{eq:sphericalDynamics}) are propagated along the horizon using a fourth order Runge-Kutta numerical integration scheme with a time step of $dt = $ 2 seconds.  The Deep RL training is performed on a laptop equipped with a six-core 2.60GHz Intel i7 CPU and an NVIDIA Quadro P3200 GPU.  

\subsection{Shuttle Vehicle Description}\label{section:NumericalExp_shuttle_descr}
We briefly describe the problem setup, the full details can be found in \cite{betts2010practical}.  The problem is concerned with a non-thrusted space reentry vehicle starting at a specified initial condition, gliding towards a specified terminal condition.  The motion of the vehicle is described by the following nonlinear dynamics,
\begin{equation}
	\label{eq:sphericalDynamics}
	\begin{aligned}
		&\dot h =v \sin\gamma , ~~~ \dot v = -\frac{D}{m}-g\sin\gamma \\
		& \dot \theta = \frac{v}{h+Re}\cos\gamma\sin\psi/\cos\phi \\
		& \dot \phi = \frac{v}{h+Re}\cos\gamma\cos\psi \\
		& \dot\gamma = \frac{L\cos\sigma}{mv} + \(\frac{v}{h+Re}-\frac{g}{v}\)\cos\gamma \\ 
		& \dot\psi = \frac{L\sin\sigma }{mv\cos\gamma}+\frac{v}{h+Re}\cos\gamma\sin\psi \frac{\sin\phi}{\cos\phi} \\
		& \dot \alpha  = \frac{1}{\tau_{\alpha}}\left[\alpha_{cmd}-\alpha \right], 
		 ~~~\dot \sigma  = \frac{1}{\tau_{\sigma}}\left[\sigma_{cmd}-\sigma \right]
	\end{aligned}
\end{equation}
where $R_e$ is radius of the Earth [m], $m$ is vehicle mass [kg], $h$ is altitude [$m$], $\theta$ is longitude [rad], $\phi$ is latitude [rad], $v$ is velocity [m/s], $\gamma$ and $\psi$ are vertical and horizontal flight path angles [rad], respectively, $\alpha$ is angle of attack [rad], and $\sigma$ is bank angle [rad].  The last two equations represent a first order dynamics model of the angle of attack and bank angle control loops, where $\tau_{\alpha}=\tau_{\sigma}=1$ second.  The following path constraints are present: $h \geq 20 \text{ [km]}$, $v \geq 600 \text{ [m/s]}$, $-20 \text{ [deg]} \leq ~\gamma \leq 20\text{ [deg]}$.
The aerodynamic and gravitational forces are computed as $L=\frac{1}{2}C_L S \rho v^2$, $D=\frac{1}{2}C_D S \rho v^2$, $\rho=\rho_0\exp(-h/H_0)$, $g=\frac{\mu}{(h+R_e)^2}$
where $S$ is surface area [m$^2$] and $H_0$, $\rho_0$, $\mu$ are Earth-specific constants. The variables $C_D=b_0+b_1\hat\alpha+b_2\hat\alpha^2$ and $C_L=a_0+a_1\hat\alpha$ are the aerodynamic drag and lift coefficients, respectively.  The state and control vectors are defined as $x=[h,v,\theta,\phi,\gamma,\psi, \alpha, \sigma]\Tr$ and $u=[\alpha_{cmd}, \sigma_{cmd}]$, respectively.  The following control constraints are imposed with clipping inside the environment: $-45 \text{ [deg]} \leq \alpha_{cmd} \leq 45\text{ [deg]}$, $-89\text{ [deg]} \leq \sigma_{cmd} \leq 89\text{ [deg]}$.

\subsection{Tracking Controller} \label{section:NumericalExp_TrackingController}
As discussed in Section \ref{section:HLAS}, a tracking controller is required in the case where $z_h(t)$ represents $\dot y^{des}(t)$.  Here we describe such a controller based on the dynamic inversion (DI) principle.  In typical DI controller synthesis the system is first linearized so the control input appears as an affine term in the system dynamics \cite{tipan2020nonlinear}.  Here we take an alternative approach which does not require any linearization.  The general idea is that for systems of the form $\dot y = a(y)+B(y)g(u)$ we can design a feedback controller by solving $B(y)g(u^*)-\left(\dot y^{des}-a(y)\right)=0$ for $u^*$ analytically in special cases \updates{(e.g. $u^*=g^{-1}\left(B(y)^{-1}\left[\dot y^{des}-a(y)\right]\right)$ assuming $B(y)$ is invertible and a closed form of the inverse function $g^{-1}(\cdot)$ can be found), or iteratively in general (e.g. via a numerical root finding method). Even though $u$ appears nonlinearly in (\ref{eq:sphericalDynamics}) as $\cos \sigma$ and $\sin\sigma$ (where bank angle $\sigma$ is a control input), we will show that $\sigma$ can be isolated using the $\arctan$ of measurable and desired quantities. The angle of attack, $\alpha$, control input is easily isolated since $C_L$ is assumed linear in $\alpha$.} 
	
We now describe the concept for the space shuttle system (\ref{eq:sphericalDynamics}).  We define the system output $y=\left[\gamma, \psi\right]\Tr$ and $\dot y^{des} = \left[\dot\gamma^{des}, \dot\psi^{des}\right]\Tr$.  
\def \fpaDot {\dot y^{des}}
\def \bankCmd {\sigma_{cmd}}
\def \alphaCmd {\alpha_{cmd}}
As a gliding vehicle, the Space Shuttle must achieve its high level goals (i.e. the flight path angle rate commands) through adjustment of its attitude with respect to the ``wind vector''. 
Since no wind is modeled, the ``wind vector'' is the velocity vector. 
The wind-relative attitude is parameterized by angle of attack $ \alphaCmd $ and bank angle $ \bankCmd $.  We will first construct an acceleration command that is a function of $ \fpaDot $. 
Assuming the vehicle has an onboard nominal aerodynamic model, the acceleration command can be used to determine the bank angle and angle of attack necessary to achieve the desired acceleration. 

\def \lat {\phi}
\def \lng {\theta}

\def \vpa {\gamma}
\def \hpa {\psi}

\def \Cev {\dcm[E][V]}

To develop this method, we first define the so-called ``velocity frame''. 
We define this frame by describing the direction cosine matrix relating the earth-fixed-inertial (ECI) frame to this velocity frame assuming a non-rotating earth (\cite{reganBook1993} Appendix H.2 and H.3), $\Cev=\dcm[N][V](\vpa,\hpa)\dcm[E][N](\lat,\lng)$.
\def \vE {\stdVec{v}[E][][E]}
By definition, the inertial velocity $ \stdVec{v}[][][E] $ of the vehicle lies completely along the x-axis of the velocity frame such that $\rowVec{v,0,0}\trans = \Cev \vE$.
%
Taking the derivative of both sides of this equation and substituting \eqref{eq:sphericalDynamics}, we can solve for the inertial acceleration of the vehicle as a function of $ \fpaDot $, $ \dot{v} $, and the current state.
\def \accelV {\stdVec{a}[V][][E]}
\def \Re {R_e}
%
\begin{align*}\label{key}
	\accelV = \begin{bmatrix}	
		\dot{v} \\
		\dfrac{\begin{matrix} v\cos(\gamma)\big(h~\dot{\psi}\cos(\phi) + \Re\dot{\psi}\cos(\phi) \\ ~~~~~~~~~~~~~~~~~~~~~~-v\cos(\gamma)\sin(\phi)\sin(\psi)\big)\end{matrix}}{\cos(\phi)(\Re + h)}\\
		- \dfrac{v\big(\Re\dot{\gamma} + \dot{\gamma}h - v\cos(\gamma)\big)}{\Re + h}	
\end{bmatrix}
\end{align*}
%
Only the force due to gravity and the aerodynamic force act on the vehicle. 
Applying Newton's second law we find that
\def \FaeroV {\stdVec{f}[V][a]}
\def \gravV {\stdVec{g}[V]}
%
%
$m \accelV	= \FaeroV + m \gravV$.
Assuming that gravity acts in the local ``down'' direction, we can define the gravity vector as $\gravV = [{-g \sin(\gamma), 0, g \cos(\gamma)}]\trans$.
Solving the above for $ \FaeroV $ yields the desired aerodynamic force that would achieve our desired acceleration.
\begin{equation}\label{key}
	\FaeroV = m(\accelV-\gravV) = \rowVec{f_1, f_2, f_3}\trans
\end{equation}
Using $ \FaeroV $ we can implement a ``bank to turn'' system. 
In ``bank to turn'', a bank angle is applied in order to line up the lift vector in the direction of desired acceleration (see \cite{reganBook1993} chapter 9.4). 
No aerodynamic side force is experienced. 
This is in contrast to a ``skid to turn'' system which uses both lift and side forces to set the aerodynamic force direction. 

The ``wind frame'' is the result of a right-handed rotation about the velocity x-axis through the bank angle. 
The lift force acts in the negative direction of the wind-frame z-axis. 
We can thus find the bank angle that ``lines up'' the lift force in the desired direction using $\bankCmd = \arctan \left({f_2}/{-f_3}\right)$.
\def \liftCmd {L_{cmd}}
Next the desired lift force magnitude $ \liftCmd $ is determined $\liftCmd=\sqrt{f_2^2 + f_3^2}$.
\def \cLift {C_{L}}
\def \sRef {S}
As stated above, the lift force is a function of the lift coefficient $ \cLift(\alpha) $, air density $ \rho $, speed $ v $, and reference surface area $ \sRef $.
We can use the equation for lift to solve $ \alphaCmd  $ such that $\liftCmd = \frac{1}{2}\cLift(\alphaCmd) \sRef \rho v^2$.  \updates{In general this may require numerical solution. In this work it can be solved analytically since $C_L$ is assumed linear in $\alpha_{cmd}$.}

\subsection{Problem 1: Maximizing Traversed Latitude}\label{section:TraversedLatitude}
We consider the following well-studied trajectory generation problem with terminal condition as defined in \cite{betts2010practical}, $h_f=24384\text{ [m]}$, $v_f =762\text{ [m/s]}$, $\gamma_f = -5\text{ [deg]}$.  
The goal is to maximize traversed latitude $\phi(t_f)$ along the horizon from a set of initial conditions.
In addition to the aforementioned path constraints, we also enforce the heating constraint $q \leq 80 \text{ [BTU/ft$^2$ - s]}$ taken from \cite{betts2010practical} at each time step along the horizon, where
$q = 779.67 (c_0 + c_1\hat\alpha + c_2\hat\alpha^2 + c_3\hat\alpha^3) \times \sqrt{\rho}(3.28084\times 10^{-4} v)^{3.07}$
is the aerodynamic heating on the vehicle wing leading edge [$\text{BTU}/\text{ft}^2-s$] and $\hat \alpha = \frac{180}{\pi} \alpha$.  Referring to Eq. (\ref{eq: J}), in this problem there is no running performance measure and the horizon reward, $\Phi$, is simply the latitude achieved at episode termination.  We follow the reward formulation described in (\ref{eq:HLAS Deep RL reward}). \updates{The instantaneous reward is set to zero, $L=0$. The horizon performance reward is designed to be positive everywhere, increasing monotonically with latitude: $\Phi = e^{\phi}\times\mathbf{1}_{\phi<0} + (1+\phi)\times\mathbf{1}_{\phi\geq0}$.  We design a ``terminal ellipsoid"
$\Psi = \left(\frac{h-h_f}{\bar h}\right)^2 + \left(\frac{v-v_f}{\bar v}\right)^2 + \left(\frac{\gamma-\gamma_f}{\bar \gamma}\right)^2$
where $\bar h=250$ [m], $\bar v=8$ [m/s] and $\bar\gamma=0.1$ [deg] are scale factors. We define the terminal reward $\hat\Psi = \min\left(1, \Psi^{-1}\right) $ which increases as the vehicle nears the center of the terminal ellipsoid. To prevent numerical issues (caused by the vehicle getting arbitrarily close to the target) we clip the terminal reward to one.} 
%
We additionally define the \textit{terminal tolerance} criteria, $|h-h_f|\leq h_{tol}$, $|v-v_f| \leq v_{tol}$, $|\gamma-\gamma_f| \leq \gamma_{tol}$, 
where $h_{tol}=500$ [m], $v_{tol}=16$ [m/s] and $\gamma_{tol}=0.5$ [deg].  The training episode is ended at the end of action-step $h$ if (i) any of the path or heating constraints are violated along the horizon, or (ii) if each of the terminal tolerance criteria is satisfied at the end of the action-step.  The Deep RL agent receives no further reward when the training episode ends.  Therefore, the agent is incentivized to satisfy all path constraints while simultaneously maximizing horizon performance reward $\Phi$ (defined above) and ending the episode inside the terminal ellipsoid level set $\Psi=1$ to maximize the terminal reward.  To promote robustness, the initial conditions are perturbed uniformly during training as $x_0=x_0^* \pm \Delta x$ where $x_0^*= [h_0,v_0,\theta_0,\phi_0,\gamma_0,\psi_0, \alpha_0, \sigma_0]\Tr$ is the nominal initial condition and $\Delta x$ is a uniformly distributed random variable with limits [4km, 390m/s, 2deg, 2deg, 2deg, 2deg, 0, 0]$\Tr$.

\subsubsection{Deep RL Comparison to NLP}\label{section:NumericalExp_DeepRLComparetoNLP}
We consider two Deep RL setups:
\begin{enumerate}
	\item The \textit{\updates{HLAS-Output}} setup is the HLAS formulation described in Section \ref{section: HLAS Deep RL Formulation}, with the ``anti-windup'' term Eq. (\ref{eq:AntiWindupLossTerm}) incorporated into the PPO surrogate objective Eq. (\ref{eq:PPOSurrogateObj}).  Each action sub-trajectory represents the desired output function $\dot y^{des}(t)=z_h(t)$.  Referring to Section \ref{section: ProblemFormulation_actionspace}, the action space is $\mathcal{A}=(\tau, \{n^{z^{(1)}}_{i}\}_{i=1}^{p+1}, \{n^{z^{(2)}}_{i}\}_{i=1}^{p+1})$, where $z^{(1)}$ corresponds to $\dot \gamma^{des}$ and $z^{(2)}$ corresponds to $\dot \psi^{des}$.  Referring to Eq. (\ref{eq:z}) we choose to represent $z$ with a linear profile, $p=1$, corresponding to a second order polynomial representation of $y^{des}$ in Eq. (\ref{eq:y^des}). \updates{The desired output function is determined completely by the choice of $z_h(t)$, and no separate trajectory optimization solution is required.  The tracking controller in section \ref{section:NumericalExp_TrackingController} ensures this output function is produced by the system.}
	\item The \textit{HLAS-Control} setup is similar to \textit{HLAS-Output}, except each action sub-trajectory represents the control input $u(t)=z_h(t)$.  Referring to Section \ref{section: ProblemFormulation_actionspace}, the action space is $\mathcal{A}=(\tau, \{n^{z^{(1)}}_{i}\}_{i=1}^{p+1}, \{n^{z^{(2)}}_{i}\}_{i=1}^{p+1})$, where $z^{(1)}$ corresponds to $\alpha_{cmd}$ and $z^{(2)}$ corresponds to $\sigma_{cmd}$, with $p=1$.  \updates{It is noted that angle of attack and bank angle are common choices for control inputs in reentry guidance problems.  However, what differentiates our work is rather than choosing these guidance angles every time step we define polynomial profiles for these angles over short (e.g. 20 second) horizons.}
\end{enumerate}

The Deep RL results are compared against a benchmark solution obtained through nonlinear programming (NLP) techniques.  To generate the NLP benchmark we utilize Pyomo \cite{hart2017pyomo}, a general, open-source, Python-based algebraic modeling language developed at Sandia National Labs.  
The final output from solving the NLP is a full trajectory along the horizon. The underlying optimization is solved through the interior-point solver \verb|IPOPT| \cite{biegler2009large}.  Details of using Pyomo in an aerospace optimal control context can be found in \cite{schlossman2021open}.

\begin{figure}
	\centering
	\begin{minipage}{0.49\textwidth}
		\includegraphics [trim = 5mm 5mm 5mm 10mm, clip, width=.99\textwidth]{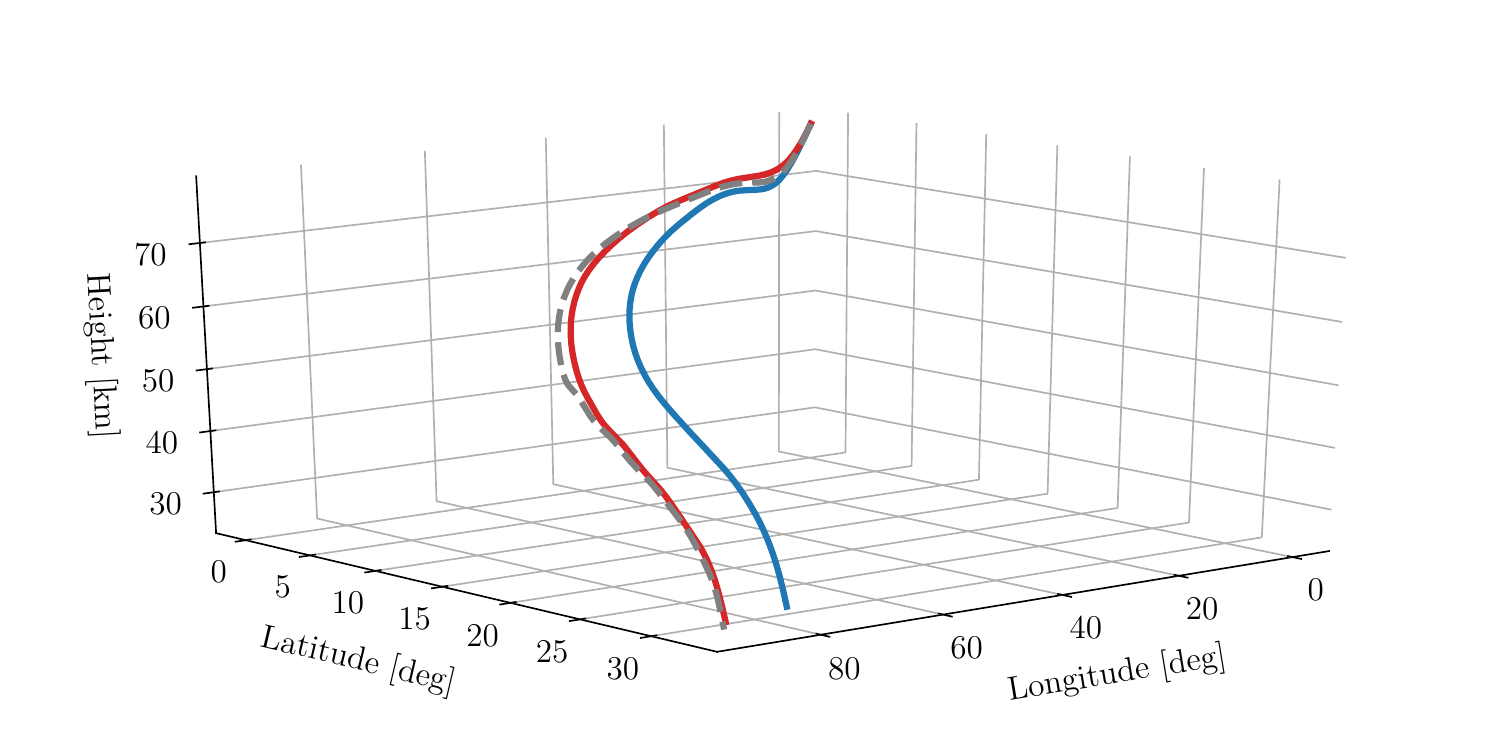}
	\end{minipage}
	\captionsetup{width=0.49\textwidth}
	\caption{Generated trajectories: Deep RL "HLAS-Output" (blue), Deep RL "HLAS-Control" (red), nonlinear programming solution (dashed grey).	
	} 
	\label{fig:shuttle_3d}
\end{figure}

\begin{figure}
	\centering
	\begin{minipage}{.49\textwidth}
		\includegraphics [trim = 0mm 0mm 5mm 0mm, clip, width=.99\textwidth]{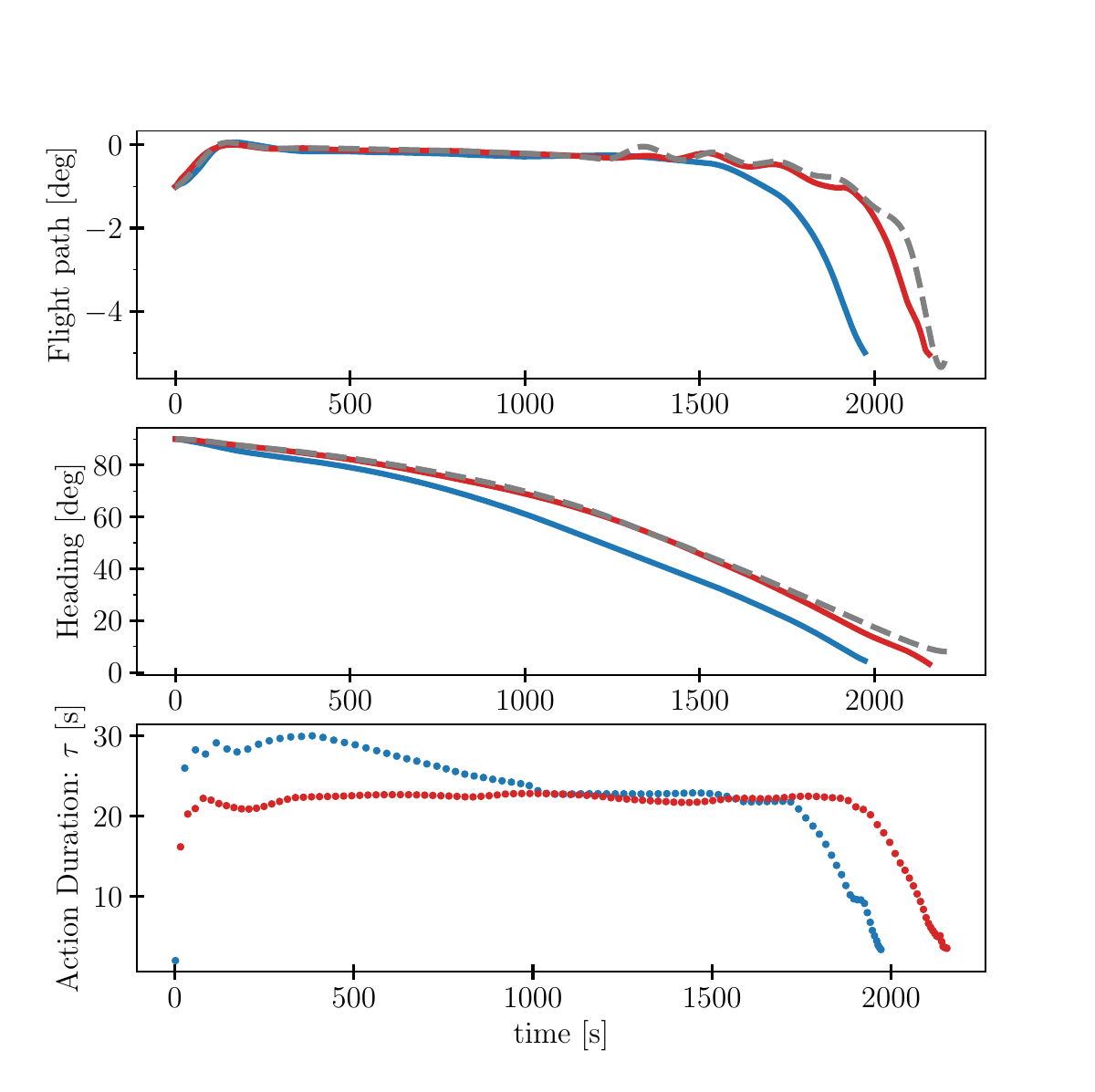}
	\end{minipage} \\
	\captionsetup{width=.49\textwidth}
	\caption{Flight angles (top, middle) and action durations (bottom). Deep RL "HLAS-Output" (blue), Deep RL "HLAS-Control" (red), nonlinear programming (dashed grey).} 
	\label{fig:shuttle_fpas}
\end{figure}

The resulting trajectory from the NLP benchmark is shown in Fig. \ref{fig:shuttle_3d}. Also shown are the Deep RL-induced trajectories, which are produced by sequentially executing the trained policy and simulated environment (see Fig. \ref{fig:HLAS_diagram}).  The Deep RL trained policy executes in approximately 0.001 seconds for arbitrary inputs to the policy network.  
The NLP solution achieves a final latitude of 32.19 [deg] and satisfies the terminal condition.  Solution times for the NLP trajectory are approximately 30 seconds.  The \textit{HLAS-Control}-trained Deep RL policy achieves a final latitude of 31.05 [deg] (96\% of the NLP solution) with terminal error $h_{\text{err}}=128.8$ [m], $v_{\text{err}}=2.14$ [m/s], $\gamma_{\text{err}}=0.035$ [deg]. The \textit{HLAS-Output}-trained Deep RL policy performs slightly worse, achieving a final latitude of 29.96 [deg].
The vertical flight path and heading angles are shown in Fig. \ref{fig:shuttle_fpas}, along with the Deep RL action duration along the horizon.  The policy tends to prefer longer action-durations throughout the horizon, except where the policy is finely adjusting the trajectory (for example, in order to maximize terminal accuracy near horizon end).

\subsubsection{Deep RL Variants}\label{section:shuttle_ablation_studies}
We now compare the performance of several Deep RL variants against the \textit{HLAS-Output} and \textit{HLAS-Control} methods.  The same network architecture (described in Section \ref{section:RLTrainingSetup}) was used throughout,  and algorithm hyperparamters were optimized when necessary.  The variants are described below:
\begin{enumerate}	
	\item The \textit{HLAS-Output-Fixed $\tau$} variant is identical to \textit{HLAS-Output} except the action duration is fixed as $\tau_h=4$ seconds.
	\item The \textit{HLAS-Output-NoAntiWindUp} variant is identical to \textit{HLAS-Output} except the ``anti-windup'' loss term Eq. (\ref{eq:AntiWindupLossTerm}) is removed from the PPO surrogate objective Eq. (\ref{eq:PPOSurrogateObj}).
	\item The \textit{Baseline} is a straightforward Deep RL setup where the attitude (angle of attack and bank angle) is commanded every timestep (two seconds).  The ``anti-windup'' loss term Eq. (\ref{eq:AntiWindupLossTerm}) is not included in the PPO surrogate objective Eq. (\ref{eq:PPOSurrogateObj}).
\end{enumerate}

\begin{figure}
	\centering
	\begin{minipage}{0.49\textwidth}
		\includegraphics [trim = 20mm 0mm 20mm 0mm, clip,width=.99\textwidth]{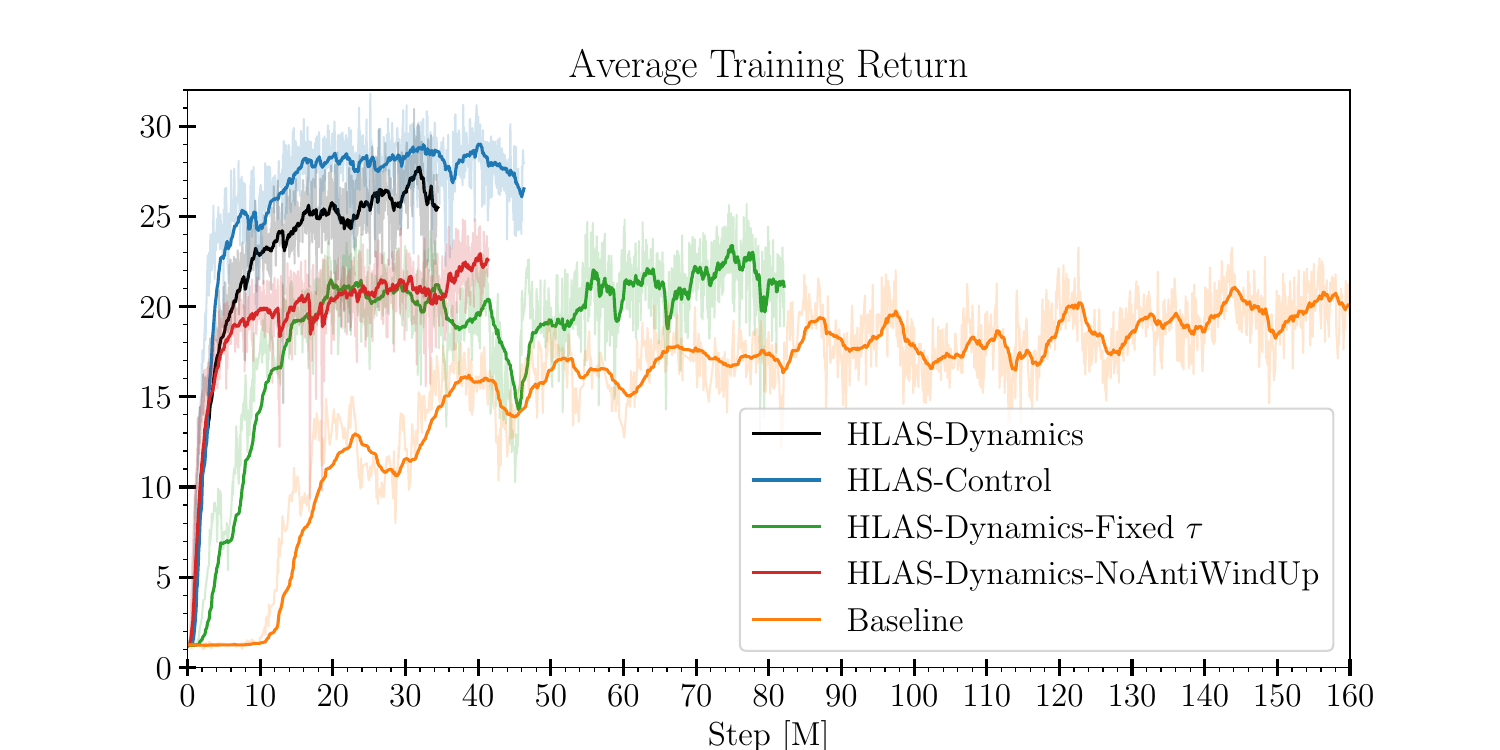}
	\end{minipage}
	\captionsetup{width=0.49\textwidth}
	\caption{Training curves for Deep RL variants.} 
	\label{fig:shuttle_baselines_training}
\end{figure}

The training curves for each variant are shown in Fig. \ref{fig:shuttle_baselines_training}.  Each training is run until a reasonable steady state is reached, up to a maximum of roughly 35 hours.  Each data point (shown as a thin line) is the average return of the previous 100 training episodes.  The thicker lines are the filtered values so the trending behavior can be easily observed.  During training a 100-episode moving average of the undiscounted episode return is maintained, and the network parameters are saved whenever a new best value is obtained.
Training hyperparameters used for each variant are shown in Table \ref{table:shuttle_baselines_hyperparameters}. \updates{The discount factor is chosen close\footnote{\updates{By convention, discount factor $\gamma$ is kept strictly less than 1.}} to 1.0 ($\gamma=0.9999$) so as not to overly discount rewards obtained at the end of long training episodes. This is especially helpful for the Baseline which requires many steps in the environment to traverse large latitude ranges (from (\ref{eq:discounted return}) the influence of a reward received $k$ steps later decays at an exponential rate of $\gamma^k$).} The entropy-based exploration bonus was included in the \textit{Baseline} and \textit{HLAS-Output-Fixed-$\tau$} variants to encourage exploration.  Various combinations of hyperparameters were used for the Baseline training, and the combination producing the best training event is shown.  Although not shown, the HLAS formulations using the PPO training algorithm was found to perform consistently well under a reasonable range of hyperparameters.
\begin{table*}[h!]
	\renewcommand{\arraystretch}{1.2}
	\caption{PPO hyperparameters. $\gamma=0.9999$, LR $=5\times 10^{-5}$ and PPO Clip $= 0.2$ in all methods.} \label{table:shuttle_baselines_hyperparameters}
	\centering 
	\begin{tabular}{lcccccc}
		\hline {Method} & $\tau_{max}, \tau_{min}$ [s] & {VF Coef} & {Ent Coef} & AntiWindup & {Steps / Proc.} & Batch Size \\
		\hline\vspace{-4mm} &&&&&& \\
		HLAS-Output & 30, 2 & 0.5 & 0 & Yes & 4096 & 128\\ 		
		HLAS-Control & 30, 2 & 0.5 & 0 & Yes & 4096 & 128\\ 
		HLAS-Output-NoAntiWindUp & 30, 2 & 0.5 & 0 & No & 4096 & 128\\ 		
		HLAS-Output-Fixed-$\tau$ & 4 [fixed] & 0.5 & 0.001 & Yes & 4096 & 128\\ 
		Baseline & 2 [fixed] & 100 & 0.001 & No & 8192 & 256\\ 
		\hline		
	\end{tabular} 	
\end{table*}

All five Deep RL methods were evaluated over 1000 randomized initial conditions  $x_0=x_0^* \pm 0.5\times\Delta x$, with $\Delta x$ as described above.  Post-training metrics are listed in Table \ref{table:shuttle_baselines_metrics}, which includes performance from the nominal initial condition, undiscounted average episode return (given by (\ref{eq:discounted return}), evaluated with discount factor $\gamma=1$), and the number of terminal misses where the state lands outside the bounds described above.  Interestingly, \textit{HLAS-Control} slightly outperforms \textit{HLAS-Output}, suggesting the policy network was sufficiently capable of representing the control dependencies within the system dynamics in this problem.  The \textit{HLAS-Output-Fixed-$\tau$} variant performs poorly, a possible explanation is the shorter action duration will see fewer states due to the shorter flight time between action samples.  The \textit{HLAS-Output-NoAntiWindUp} variant also performed poorly. Although it is not shown here, with \textit{HLAS-Output-NoAntiWindUp} the action mean drifts beyond the action limit in several parts of the trajectory during training and never falls back within limits.
\begin{table*}[ht]
	\renewcommand{\arraystretch}{1.2}
	\caption{Post training metrics.  Averages were taken over 1000 initial condition perturbations.} \label{table:shuttle_baselines_metrics}
	\centering 
	\begin{tabular}{lp{25mm}p{27mm}p{25mm}p{25mm}}
		\hline {Method} & Latitude Achieved\newline $\Phi$ (nominal) & Terminal Miss Dist. \newline $\Psi$ (nominal) &  Undiscounted \newline Ep. Ret. (average) & \# Terminal Misses \newline (out of 1000) \\
		\hline\vspace{-4mm} & & & & \\
		HLAS-Output & 29.96 deg & 0.305 & 34.67 & 0 \\ 		
		HLAS-Control & 31.05 deg & 0.679 & 35.26 & 0 \\ 
		HLAS-Fixed-$\tau$ & 27.96 deg & 27.78 & 28.94 & 955 \\ 
		HLAS-NoAntiWindUp & 24.72 deg & 2.08 & 28.65 & 7\\ 		
		Baseline & 26.39 deg & 1.67 & 29.38 & 21\\ 
		\hline		
	\end{tabular} 	
\end{table*}

\subsection{Problem 2: Trajectory Planning for Debris Avoidance}\label{section:DebrisAvoidance}
In the previous section we showed our Deep RL approach can solve a difficult trajectory generation problem which is well-suited for traditional optimal control solvers (e.g. nonlinear programming).  We now demonstrate the effectiveness of our approach as a trajectory planner on a quasi-static map, and show that traditional optimal control solvers struggle here.  In this problem we employ the \textit{HLAS-Control} variant and again consider the shuttle vehicle described in Section \ref{section:NumericalExp_shuttle_descr}.  The objective is to guide the shuttle path to reach a specific reentry location $h_f=24384\pm1000$[m], $\theta_f=0\pm 1$ [deg], $\phi_f=50\pm 1$ [deg] while avoiding floating space debris represented as obstacles at fixed locations.  In Fig. \ref{fig:shuttle_obstacle_field_deeprl_results} the terminal location is shown as the blue dot and the debris field is shown as the collection of light red ellipses.  
The initial shuttle location is randomly set along a semi-circle in the longitude-latitude plane of radius = $50 \pm 5$ [deg] surrounding the terminal location.  To ensure the problem is feasible the initial heading, $\psi_0$, points the shuttle vehicle towards the terminal location with a randomly set $\pm 5$ degree error.  Other initial conditions are $h_0=79248 \pm 2000$ [m], $v_0=7802 \pm 100$ [m/s], $\gamma=-1\pm 1$ [deg].  

For Deep RL training we again use the reward structure described in (\ref{eq:HLAS Deep RL reward}): $L=0$, $\Phi = 0$,  
$\Psi = \left(\frac{h-h_f}{\bar h}\right)^2 
		+ \left(\frac{\theta-\theta_f}{\bar \theta}\right)^2 
		+ \left(\frac{\phi-\phi_f}{\bar \phi}\right)^2 $
with terminal reward $\hat\Psi = \min\left(1, \Psi^{-1}\right)$ and $C_0 = 5$. Scale factors are set as $\bar h=1000$ [m], $\bar \theta=1$ [deg] and $\bar\phi=1$ [deg].  In this problem the performance reward $\Phi$ is zero, since the goal is to reach the target location without violating a state constraint or crossing an obstacle.   Each training episode (successfully) ends when three simultaneous terminal conditions are satisfied: $|h-h_f|\leq 500$ [m], $|\theta-\theta_f|\leq 1$ [deg], $|\phi-\phi_f| \leq 1$ [deg].  The training episode ends prematurely if the agent intersects any of the obstacles shown in Fig. \ref{fig:shuttle_obstacle_field_deeprl_results}.  Over approximately 16 hours of off-line training the agent learns to avoid the obstacles in order to maximize the terminal reward.  We train with the \textit{HLAS-Control} variant described in the previous section.  Referring to Section \ref{section: ProblemFormulation_actionspace}, the action space is $\mathcal{A}=(\tau, \{n^{z^{(1)}}_{i}\}_{i=1}^{p+1}, \{n^{z^{(2)}}_{i}\}_{i=1}^{p+1})$, where $z^{(1)}$ corresponds to $\alpha_{cmd}$ and $z^{(2)}$ corresponds to $\sigma_{cmd}$.  We choose $p=1$ corresponding to linear profiles for $\alpha_{cmd}$ and $\sigma_{cmd}$.  

\subsubsection{Trajectory Planning Results}
Planning with a trained policy is straight-forward, which is a significant benefit of using Deep RL: trajectories are produced by sequentially executing the trained policy and then evaluating in the simulated environment, as shown in Fig. \ref{fig:HLAS_diagram} and Algorithm 2.  \updates{The action duration, $\tau$, and polynomial coefficients for angle of attack, $poly_\alpha$, and bank angle, $poly_\sigma$, are extracted from the policy network's action-mean $\mu_\theta(s)$ as shown in line 2. In lines 5 and 6 commands for angle of attack and bank angle are generated by evaluating each polynomial at simulation time $t$ with} \verb|EvalPoly|.  The \verb|Simulator|\footnote{In real application, the simulator would be replaced with the actual vehicle and a new state would be measured every $dt$ seconds.} \updates{numerically propagates the shuttle vehicle dynamics Eq. (\ref{eq:sphericalDynamics}) with a numerical timestep of $dt$. After $\tau$ seconds of simulation, the updated state $s$ is passed to the policy network's action-mean and a new action is computed.}  
The planning process ends when the shuttle reaches its goal state or intersects a debris obstacle.  In this work the debris field and goal state are quasi-static, so these pieces of information are not provided to the policy (the policy network has essentially memorized this particular map configuration).  Given that end-to-end training completes in 16 hours on our laptop hardware, this is not an unreasonable assumption depending on the application.  However, in future work we seek to perform planning in response to dynamic obstacle fields and goal states, in which obstacles and goal states are provided as inputs to the policy, so that the policy can learn to generalize to arbitrary maps.    

\begin{algorithm}[] 
	\SetAlgoLined
	\KwIn{Trained policy $\mu_\theta$, initial state $s$, simulator timestep $dt$}
	\While{done is False}{
		$[\tau, poly_{\alpha}, poly_{\sigma}] = \mu_\theta(s)$ \# extract action\\
		initialize: $t=0$\\
		\While{$t\leq \tau$}{
			$\alpha_{cmd} = \text{EvalPoly}(poly_\alpha, t)$\\
			$\sigma_{cmd} = \text{EvalPoly}(poly_\sigma, t)$\\
			$s\leftarrow$ Simulator($\alpha_{cmd}, \sigma_{cmd}, s, dt$) \\
			$t\leftarrow t+dt$\\
			\eIf{state limit violation \Or hit debris obstacle \Or reach goal}{
				done = True\\
			}{
				done = False
			}
		}
	}
	\caption{Trajectory planning with Deep RL policy}
\end{algorithm}

\begin{figure}
	\centering
	\begin{minipage}{0.50\textwidth}
		\includegraphics [trim = 20mm 0mm 20mm 0mm, clip,width=.99\textwidth]{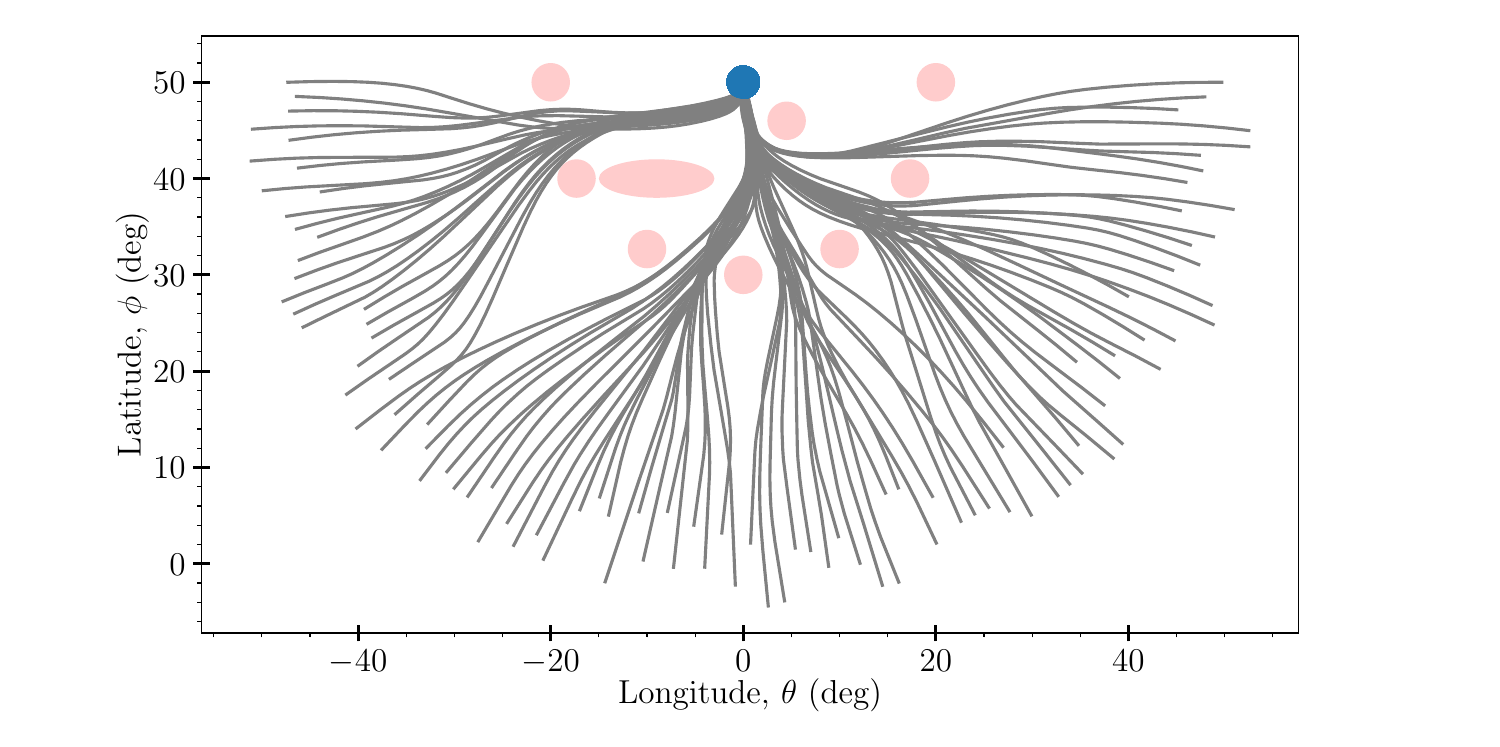}
	\end{minipage}
	\captionsetup{width=.49\textwidth}
	\caption{Deep RL planner results from 100 randomly chosen initial conditions.  Floating debris field (light red ellipses) and reentry target location (blue dot).} 
	\label{fig:shuttle_obstacle_field_deeprl_results}
\end{figure}

Trajectory planning results for the Deep RL planner for 100 initial conditions distributed over the semi-circle radius are shown in Fig. \ref{fig:shuttle_obstacle_field_deeprl_results}.  The initial conditions have been perturbed about nominal values as discussed above.  
In each case the Deep RL planner is able to guide the shuttle to its reentry location within tolerance without intersecting any of the obstacles.  This problem was also attempted with the NLP methods discussed in Section \ref{section:NumericalExp_DeepRLComparetoNLP}.  Our implementation of NLP had difficulty solving this problem consistently, and required careful trial-and-error calibration of the numerical discretization scheme.  If the number of discretization points was too large or too small the solver would not converge.  Additionally, a set of discretization points which worked for one initial condition did not necessarily work for another.  This can be problematic for real-time planning as each NLP solution can take several minutes or longer if the solver gets ``stuck" while trying to find a solution.  

\begin{figure}
	\centering
	\begin{minipage}{0.55\textwidth}
		\includegraphics [trim = 25mm 0mm 15mm 0mm, clip,width=.99\textwidth]{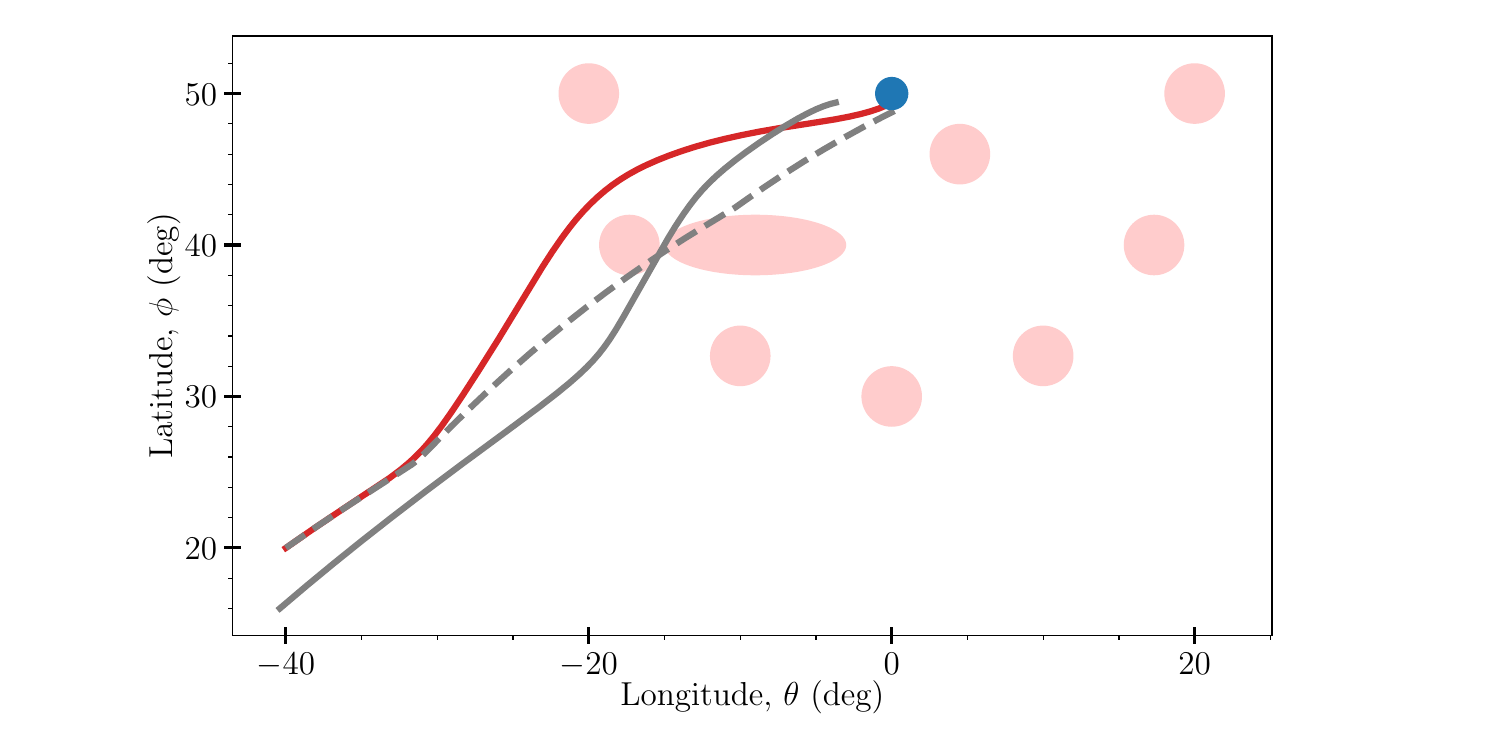}
	\end{minipage}
	\captionsetup{width=.49\textwidth}
	\caption{Deep RL (red) comparison to NLP (grey) on a challenging initial condition.} 
	\label{fig:shuttle_obstacle_field_results_trappedsoln}
\end{figure}

In an attempt to promote convergence robustness with the NLP solver we also attempted a ``warm-start" procedure where the problem was first solved without obstacles present, and then re-solved with obstacles present using the first solution as an initial guess for the solver.  This often worked, but at times the trajectory became ``trapped" between two obstacles as shown in Fig. \ref{fig:shuttle_obstacle_field_results_trappedsoln}.  The problem is first solved by NLP with the debris field absent (dashed grey), this solution is provided as an initial guess to the NLP solver in a second iteration where the debris field is present.  On this second iteration the NLP trajectory (solid grey) becomes ``trapped" between two obstacles and cannot satisfy the boundary conditions.

\section{Conclusions}
This paper presents an approach for trajectory planning based on Deep RL, where actions are sub-trajectories of variable duration and shape.  The developed approach is referred to as the high-level action space approach.  The HLAS approach appears to promote exploration within the environment as shown through a brief analysis and demonstrated empirically.  The HLAS approach is shown to significantly improve the Deep RL training process on a long-range trajectory generation problem, and also outperforms NLP on a trajectory planning problem involving obstacle avoidance.

We first assessed our HLAS approach on a well-studied long-range trajectory generation problem using the shuttle reentry vehicle.  First, off-line training is performed over a range of initial conditions to produce a state-feedback policy (policy training).  Second, after training is complete, the policy is rapidly evaluated in response to state inputs to induce a trajectory in a sequential manner (trajectory generation).  During the trajectory generation phase of Deep RL, the trained policy achieves optimality similar to NLP solutions, achieving 96\% of the NLP solution objective.  At trajectory generation time, the trained Deep RL policy can be evaluated rapidly making it a good candidate for real-time implementation.  The Deep RL policy can be evaluated in a few milliseconds, whereas the full NLP solution takes approximately 30 seconds to obtain on the same hardware.  
On the same problem our approach outperforms a straight-forward implementation of Deep RL where actions are low level inputs, and action decisions are made every time step (in contrast to the HLAS approach where a constrained input sequence executes over multiple time steps).  The trained policy from HLAS Deep RL produces an 18\% improvement in the objective function, more efficiently and with less sensitivity to hyperparameters, compared to the straight-forward implementation of Deep RL.  The HLAS method achieves steady state during training with approximately 75\% fewer training steps than standard approaches.

Finally, we investigated shuttle trajectory planning around a fixed obstacle field and fixed goal state.  Training is performed off-line using the HLAS Deep RL approach, and the trained policy is used directly as the planner from randomized initial conditions to the fixed goal state.  Our planning approach is shown to produce solutions more robustly and more rapidly than NLP. Our method improves planning speed and stability by moving the computational burden off-line, where difficult scenarios are worked out through randomized exploration of the state-space.  Total training time is approximately 16 hours on our hardware for this problem.  In future work we seek to augment the HLAS Deep RL training, allowing the policy to plan in response to arbitrarily placed obstacles and goal states.

\section*{Funding Statement}
This work was supported by the Laboratory Directed Research and Development program at Sandia National Laboratories, a multi-mission laboratory managed and operated by the National Technology and Engineering Solutions of Sandia LLC, a wholly owned subsidiary of Honeywell International Inc. for the U.S. Department of Energy’s National Nuclear Security Administration under contract DE-NA0003525. This paper describes objective technical results and analysis. Any subjective views or opinions that might be expressed in the paper do not necessarily represent the views of the U.S. Department of Energy or the United States Government.

\bibliographystyle{ieeetr}
\bibliography{paper_v1}

\end{document}